\documentclass[12pt,preprint]{aastex}
\usepackage{subfig}

\begin{document}
\title{Contemporaneous VLBA 5 GHz Observations of LAT-Detected Blazars}
\author{J. D. Linford\altaffilmark{1}, G. B. Taylor\altaffilmark{1}, R. W. Romani\altaffilmark{2},
J. F. Helmboldt\altaffilmark{3},  A. C. S. Readhead\altaffilmark{4}, R. Reeves\altaffilmark{4}, and J. L. Richards\altaffilmark{4}}

\altaffiltext{1}{Department of Physics and Astronomy, University of New Mexico, MSC07 4220, Albuquerque, NM 87131-0001}
\altaffiltext{2}{Department of Physics, Stanford University, Stanford, CA 94305}
\altaffiltext{3}{Naval Research Laboratory, Code 7213, 4555 Overlook Ave. SW, Washington, DC 20375}
\altaffiltext{4}{Astronomy Department, California Institute of Technology, Mail Code 247-17, 1200 East California Boulevard, Pasadena, CA 91125}

\begin{abstract}

The radio properties of blazars detected by the Large Area Telescope (LAT) on board the \textit{Fermi Gamma-ray Space Telescope} have been observed contemporaneously by the Very Long Baseline Array (VLBA).  In total, 232 sources were observed with the VLBA.  Ninety sources that were previously observed as part of the VLBA Imaging and Polarimetry Survey (VIPS) have been included in the sample, as well as 142 sources not found in VIPS.  This very large, 5 GHz flux-limited sample of active galactic nuclei (AGN) provides insights into the mechanism that produces strong $\gamma$-ray emission.  In particular, we see that $\gamma$-ray emission is related to strong, uniform magnetic fields in the cores of the host AGN.  Included in this sample are non-blazar AGN such as 3C84, M82, and NGC 6251.  For the blazars, the total VLBA radio flux density at 5 GHz correlates strongly with $\gamma$-ray flux.  The LAT BL Lac objects tend to be similar to the non-LAT BL Lac objects, but the LAT flat-spectrum radio quasars (FSRQs) are significantly different from the non-LAT FSRQs.  Strong core polarization is significantly more common among the LAT sources, and core fractional polarization appears to increase during LAT detection.

\end{abstract}


\section{Introduction}

The Large Area Telescope (LAT; Atwood et al.\ 2009) on board the \textit{Fermi Gamma-ray Space Telescope} is a wide-field telescope covering the energy range from about 20 MeV to more than 300 GeV.  It has been scanning the entire $\gamma$-ray sky once every three hours since July of 2008.  The LAT first-year catalog (1FGL; Abdo et al.\ 2010a) has 1451 sources.  The majority of these $\gamma$-ray bright sources that have been identified with radio sources are associated with blazars (685 of 1451).  These blazars typically are strong, compact radio sources which exhibit flat radio spectra, rapid variability, compact cores with one-sided parsec-scale jets, and superluminal motion in the jets (Marscher 2006).

Several very long baseline interferometry (VLBI) programs, such as the Monitoring Of Jets in AGN with VLBA Experiments (MOJAVE; Lister et al.\ 2009a; Homan et al.\ 2009; Lister et al. 2011) observing at 15 GHz, the Boston University program observing at 22 and 43 GHz (Marscher et al.\ 2010; Jorstad et al.\ 2010), and TANAMI (Ojha et al.\ 2010a) observing at 8.4 and 22 GHz, along with the LAT collaboration themselves (Abdo et al. 2011) are in place to monitor the radio jets from the brightest blazars such as 3C273, BL Lac, etc.  Our sample has a flux limit roughly an order of magnitude below the MOJAVE survey and so allows us to probe the extensions of the radio core/$\gamma$-ray properties down to a fainter population.  Many different radio-$\gamma$-ray correlations have been suggested (Taylor et al.\ 2007, Abdo et al.\ 2009a, Lister et al.\ 2009a, Kovalev et al.\ 2009, Giroletti et al.\ 2010, Ojha et al. 2010b, Linford et al. 2011).  In our last paper (Linford et al. 2011) we found only a marginal correlation between $\gamma$-ray flux and radio flux density, and only for the FSRQs.  With our new, more contemporaneous and larger sample, we find a strong correlation between the two.  By examining a larger sample (1248 objects) we attempt to obtain more definitive insight into several other properties of $\gamma$-ray bright blazars.

In Section 2 we define our sample.  In Section 3 we briefly describe the data reduction process for our VLBA observations.  In Section 4 we present data on the $\gamma$-ray flux and radio flux densities of the LAT sources.  In Section 5 we discuss differences between sources observed in two epochs (prior to or during 2006 and in 2009 - 2010). In Section 6 we compare several parameters of the LAT and non-LAT sources. Notes on some individual objects of interest are presented in Section 7 and in Section 8 we discuss our results.  Throughout this paper we assume $H_{0} = 71$ km s$^{-1}$ Mpc$^{-1}$ and $\Lambda$CDM cosmology (e.g., Hinshaw et al.\ 2009).

\section{Sample Definition}

We obtained time on the Very Long Baseline Array (VLBA) to observe LAT-detected sources from November of 2009 to July of 2010.  We had a total of 7 observing runs and collected 5 GHz data on 232 sources.  The first 3 observing runs were follow-up observations on 90 sources in the VLBA Imaging and Polarimetry Survey (VIPS; Helmboldt et al.\ 2007) and new 5 GHz observations of 7 sources in the MOJAVE sample.  The remaining 135 sources were selected from the Fermi 1FGL as sources which were associated with a source in the Combined Radio All-Sky Targeted Eight GHz Survey (CRATES; Healey et al. 2007) with high ($\geq$80\%) probability, had a flux density of at least 30 mJy in CRATES, and were not in VIPS or MOJAVE.  Of the 232 sources, 95 are BL Lac objects, 107 are FSRQs, and 30 are other types of AGN (radio galaxies, AGN of unknown type, and 1 starburst galaxy).  Any object that is not a BL Lac object or FSRQ we classify as `AGN/Other'.  The optical classifications are taken from the LAT LAT First Catalog of AGN (1LAC; Abdo et al. 2010b).  See Table~\ref{datatable} for a summary of our data.  

We should note that several of the 1FGL sources are associated with multiple AGN in the 1LAC.  Not all of the associated sources are in CRATES, and not all of them are brighter than 30 mJy at 8 GHz.  We make a note of those objects with multiple associations in Table~\ref{datatable}.  We observed 2 radio sources for 3 of the LAT sources in our sample.  The LAT source 1FGL J1225.8+4336 is associated with the BL Lac object VIPS J12248+4335 (probability 94\%) and the FSRQ VIPS J12269+4340 (probability 87\%).  The LAT source 1FGL J0448.6+1118 is associated with the FSRQ CRATES J0448+1127 (F04486+112A in our sample; probability 99\%) and the BL Lac object PKS 0446+11 (F04486+112B in our sample; probability 92\%).  The LAT source 1FGL J0510.0+1800 is associated with an AGN of unknown type CRATES J0509+1806 (F05100+180A; probability 91\%) and the FSRQ PKS 0507+17 (F05100+180B; probability 100\%).  It is possible that all of these sources are emitting $\gamma$-rays, so we included all of them in our calculations.  Where possible, we use the 1LAC redshifts.  If the source did not have a redshift listed in 1LAC, we searched the NASA/IPAC Extragalactic Database (NED).

To build a non-LAT detected sample for comparison, we excluded all LAT sources from VIPS, leaving 1018 objects in our non-LAT sample.  Of these 1018 objects, 24 are BL Lacs, 479 are FSRQs, and 515 are AGN/Other types (radio galaxies or AGN of uncertain type).  The optical types for VIPS sources were adopted from the Candidate Gamma-Ray Blazar Survey (CGRaBS; Healey et al. 2008).  

\section{VLBA Data Reduction}

All VLBA data were correlated using the new DiFX software correlator (e.g., Deller et al. 2011).  The correlated data was processed via the VIPS data reduction pipeline using automated scripts.  See Taylor et al. (2005) and Helmboldt et al. (2007) for a more thorough description of the VIPS pipeline.  Initial calibration (fringe-fitting, flux and phase calibration, polarization calibration) was done using the Astronomical Image Processing System (AIPS; Greisen 2003).  Imaging and visibility model-fitting were done using Difmap (Shepherd 1997).

\section{Gamma-ray Flux and Radio Flux Density}

Throughout the rest of this paper, we will use the nonparametric Spearman test (e.g., Press et al.\ 1986) to look for correlations between the LAT-detected and non-LAT-detected objects.  The Spearman correlation coefficient ($\rho_{s}$) has a range of $0<|\rho_{s}|<1$.  A high value of $\rho_{s}$ indicates a significant correlation.  This is a powerful test for statistical correlation, but it does not test an actual physical correlation.  In order to be certain our correlations are physically significant, we ensure that a redshift selection effect is not adding a bias to our data.  

The LAT measures $\gamma$-ray flux in several bands.  To create total $\gamma$-ray fluxes, we combined the fluxes from 3 bands: 100-300 MeV, 300 MeV - 1 GeV, and 1-100 GeV.  The fluxes were added and uncertainties were added in quadrature.  However, some sources had only upper limits to their fluxes in certain bands.  If a source's reported fluxes in one or two bands were upper limits, we use 1/2 the reported flux as the uncertainty in that band as the upper limits are given as 2-sigma results (Abdo et al. 2010a).  If a source had upper limits on its flux in all three bands, we adopt the convention of listing its error as 0.00.

\subsection{Redshift Selection Effect}

Imagine a population of sources covering a wide range of redshifts in which the radio and $\gamma$-ray emission is not physically correlated, but in which radio flux and $\gamma$-ray flux are correlated with redshift.  This is actually what one would naively expect because the more distant objects will be fainter.  Such a population will show a strong correlation of radio flux with gamma ray flux via the Spearman test, but this does not indicate a significant physical correlation between these two observables.  

However, by investigating the relationship between radio flux density at 5 GHz ($S_{5}$) and redshift for the LAT-detected sources we can rule out a correlation between the two.  We calculated the Spearman $\rho_{s}$ values for the $S_{5}$-z relationship for the BL Lac objects and FSRQs separately.  For the BL Lacs, the $\rho_{s}$ value was 0.308, with a 2.1\% probability that random sampling would produce this same $\rho_{s}$ value.  For the FSRQs, the $\rho_{s}$ value was 0.109, with a 26.2\% probability of getting the same value by random sampling.  Therefore, there is a very marginal correlation between the radio flux density and redshift for our BL Lac objects, and no correlation for our FSRQs.

We also tested the correlation between $\gamma$-ray flux and redshift.  The $\rho_{s}$ values were 0.084 for BL Lacs and 0.016 for FSRQs, with the probability of getting the same values from random sampling of 54.0\% and 86.8\%, respectively.  So, there is no significant correlation between $\gamma$-ray flux and redshift for either BL Lac objects or FSRQs.  We also visually inspected plots of both LAT flux and total VLBA radio flux density versus redshift and saw no obvious trends, confirming the correlation coefficient results. 

\subsection{Gamma-ray Flux versus Radio Flux Density}

\begin{figure}
\plotone{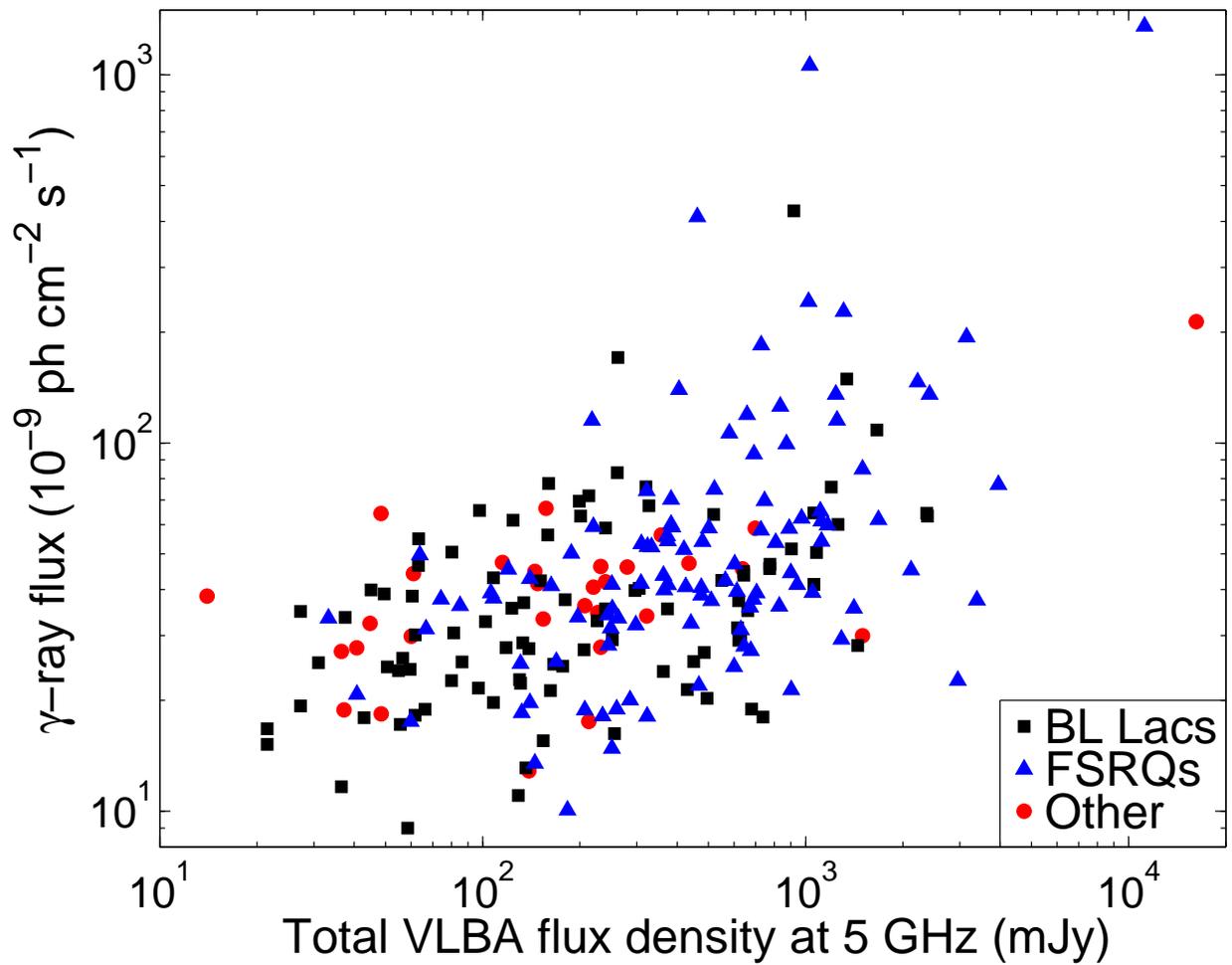}
\caption{LAT $\gamma$-ray flux (100 MeV - 100 GeV) vs. total VLBA radio flux density at 5 GHz.  The $\gamma$-ray fluxes are in units of $10^{-9}$ photons cm$^{-2}$ s$^{-1}$.  The black squares are BL Lacs, the blue triangles are FSRQs, and the red circles are radio galaxies and unclassified objects.  Error bars are omitted for ease of viewing.}
\label{CombFluxFlux}
\end{figure}

In Fig.~\ref{CombFluxFlux} we plot the LAT flux versus the total VLBA flux density at 5 GHz.  Again, the LAT fluxes are broadband fluxes from 100 MeV to 100 GeV.  

The Spearman $\rho_{s}$ values we found were 0.467 for the BL Lacs and 0.510 for the FSRQs.  The probabilities for getting the same $\rho_{s}$ values from random sampling were 2$\times$10$^{-6}$ for the BL Lac objects and 2$\times$10$^{-8}$ for the FSRQs.  Therefore, the LAT flux correlates very strongly with the total VLBA flux density.  The BL Lac objects do not correlate quite as strongly, especially considering that we found a marginal correlation between total VLBA flux density and redshift for these objects in the previous section.  Still, this provides solid evidence that objects with higher radio flux density also produce more $\gamma$-ray flux.

In our previous paper (Linford et al. 2011), we found that there was no strong correlation between radio flux density and $\gamma$-ray flux.  However, our flux densities for those observations were prior to the launch of \textit{Fermi}.  Furthermore, it is likely that the radio flux density increases during episodes of $\gamma$-ray flaring (e.g., Kovalev et al. 2009).  Our newer, larger, and more contemporaneous set of observations are more appropriate for making these comparisons.

\subsection{Radio Flux Density}

\begin{figure}
\plotone{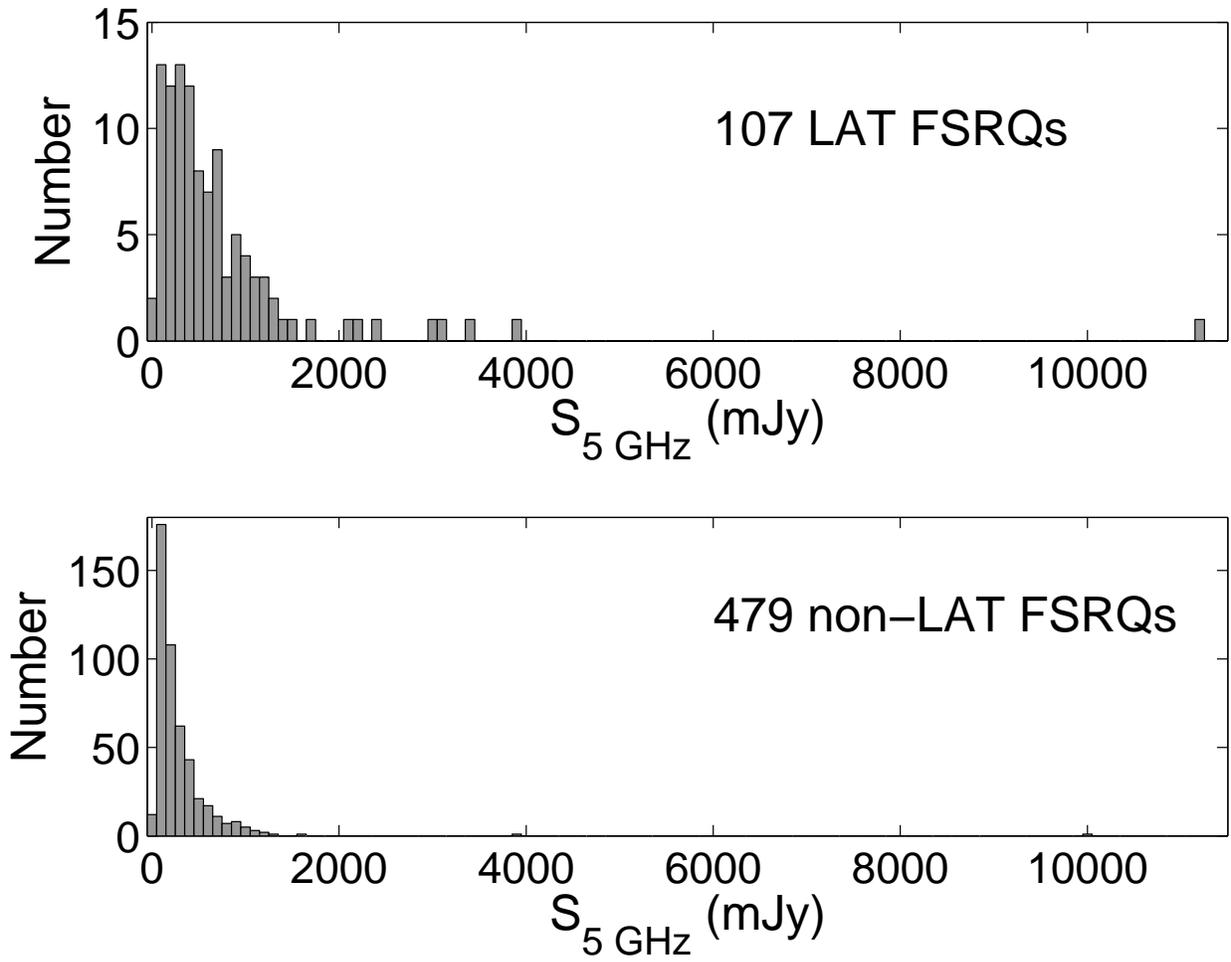}
\caption{Distributions of LAT (top) and non-LAT (bottom) FSRQ total VLBA flux density at 5 GHz.}
\label{fsrq_s5}
\end{figure}

The median total VLBA flux density at 5 GHz was 177 mJy for the LAT BL Lacs and 221 mJy for the non-LAT BL Lacs.  The LAT FSRQ appeared to have significantly higher flux densities than the non-LAT FSRQ, with a median of 467 mJy for LAT FSRQs compared to a median of only 191 mJy for non-LAT FSRQs.  The LAT AGN/Other objects also had higher 5 GHz flux densities than their non-LAT sources counterparts.  
The Kolmogorov- Smirnov (K-S) test probability that the LAT and non-LAT BL Lac objects belong to the same parent population is about 38\%.  The K-S test probability that the LAT and non-LAT FSRQs are related is only 3$\times$10$^{-12}$.  See Fig.~\ref{fsrq_s5} for a plot of the FSRQ total VLBA 5 GHz flux density distributions.

\section{Archival versus Contemporaneous}
In this section we investigate the changes in the 90 sources that were part of VIPS (i.e., observed in 2006 or earlier) and our new sample (observed in 2009-2010).  Of these 90 sources, 35 are BL Lac objects, 44 are FSRQs, and 11 are AGN/other objects.  It is important to note that we do not know whether most of our sources were emitting $\gamma$-rays in the archival data.  It is possible that many of them would have been detected by \textit{Fermi} had it been in operation at the time.  Only 6 of these sources were detected by the Energetic Gamma Ray Experiment Telescope (EGRET) on board the \textit{Compton Gamma-ray Observatory} satellite.  As an estimate of the uncertainty of the median values we will use the median absolute deviation (MAD) given by
\begin{equation}
MAD(X) = median(|(X_{i}-median(X)|)
\end{equation}
where $X_{i}$ is a value in the array $X$.  In other words, it is the median of the positive difference between each entry in X and the overall median of X.  See Huber \& Ronchetti (2009) for more information on the MAD.

\subsection{Radio Flux Density Variability}

\begin{figure}
\plotone{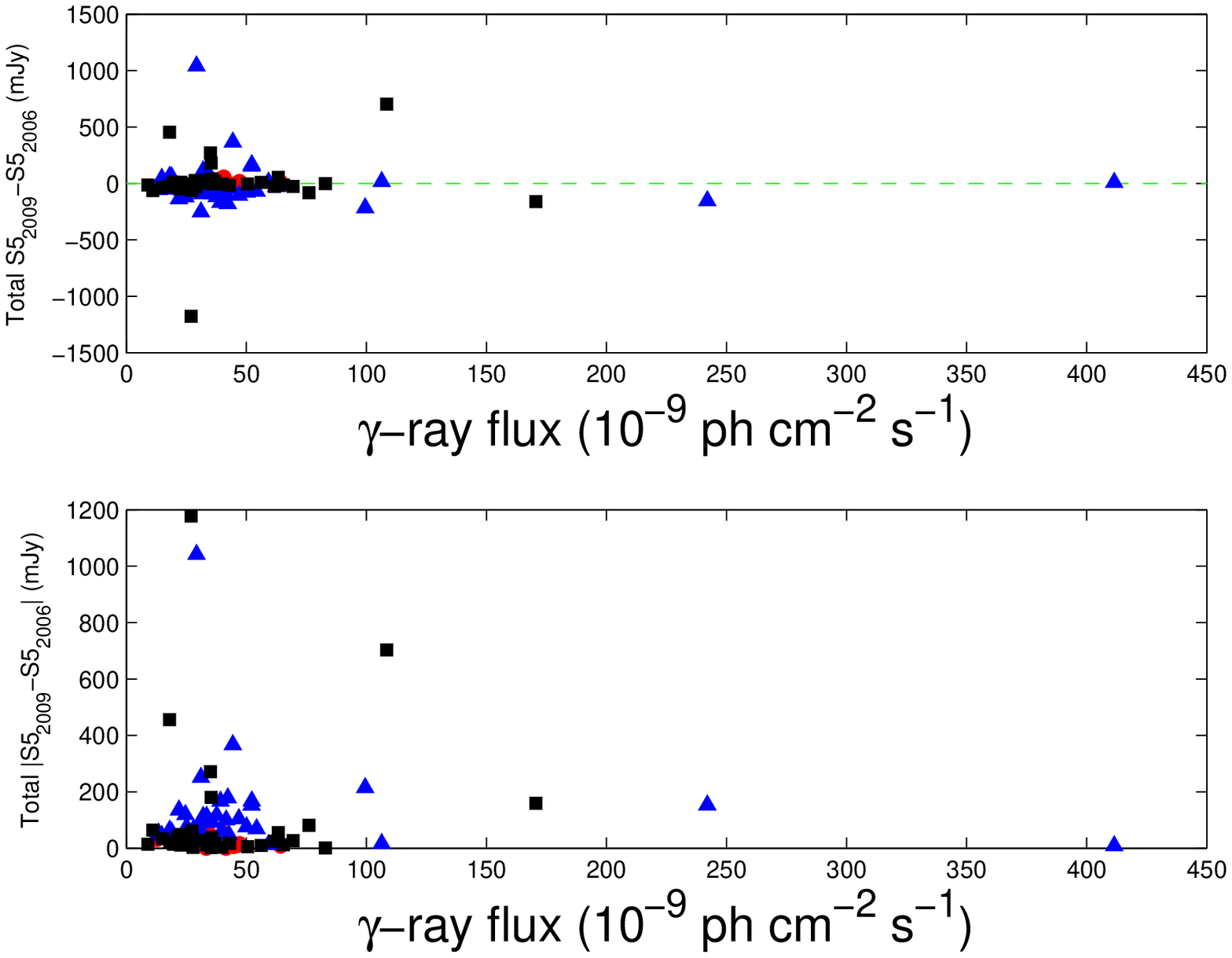}
\caption{Top: Difference in total flux density at 5 GHz (current - archival) versus total LAT flux.  Bottom: Absolute value of the difference in total flux density (current - archival) versus total LAT flux.  Black squares are BL Lac objects, blue triangles are FSRQs, and red circles are AGN/other.}
\label{tS5diff}
\end{figure}

It is well known that blazars are highly variable sources in the radio and $\gamma$-ray regimes. Several studies have also indicated that the radio and $\gamma$-ray variability are correlated (e.g., Kovalev et al. 2009 and Schinzel et al. 2010). Looking at the differences in total VLBA radio flux densities (see Fig.~\ref{tS5diff}) for 90 of our sources that were observed both in 2006 (or earlier) and in 2009, we saw that while there is variability the magnitude of the variability does not appear to correlate with the $\gamma$-ray flux.  The median total flux densities showed little significant change between the two epochs.  The BL Lac objects had a median flux density of 149.2 mJy in the archival data and 133.2 mJy in the new data.  The difference between the medians for the two epochs falls well within the median absolute deviations of 74 mJy for archival data and 72 mJy for the current data.  The FSRQs appeared to become somewhat brighter in the new data, with a median flux density of 256.2 mJy in the archival data and 316.0 mJy in the new data.  Again, the difference between the two epochs is not significant when compared to the median absolute deviations of 122 mJy for the archival data and 176 mJy for the current data.

\begin{figure}
\plotone{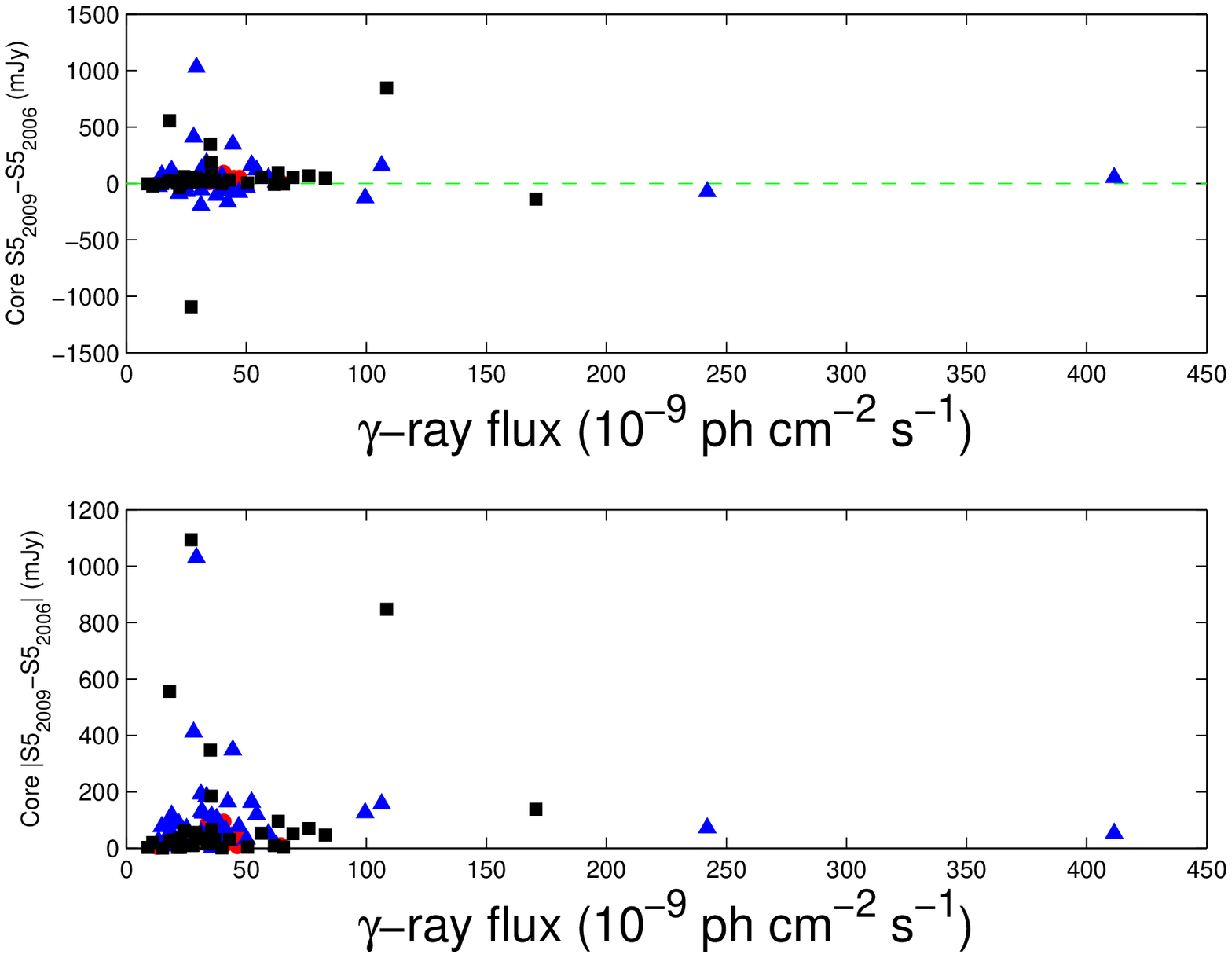}
\caption{Top: Difference in core flux density at 5 GHz (current - archival) versus total LAT flux.  Bottom: Absolute value of the difference in core flux density (current - archival) versus total LAT flux.  Black squares are BL Lac objects, blue triangles are FSRQs, and red circles are AGN/other.}
\label{cS5diff}
\end{figure}

We also looked at the radio emission from the core (the bright compact component at the base of the jet). The core flux density was found using automatic visibility fitting (see Taylor et al. 2007 for a more detailed discussion of this process). See Fig.~\ref{cS5diff} for a plot of the difference in core flux density (current minus archival) versus the total LAT flux.  Again, we found that there is no apparent correlation between core flux density variability with LAT flux. The median core flux for the FSRQs is higher during $\gamma$-ray detection, going from 225.1 mJy in the archival data to 293.6 mJy in the new data, but this is again within the median absolute deviations of 107 mJy for the archival data and 156 mJy for the new data.  Also, from Fig.~\ref{cS5diff} we can see that most of the FSRQs are actually slightly dimmer, with 25 of the 44 FSRQs showing a reduction in in core flux density.  

\subsection{Core Brightness Temperature Variability}
We obtained the brightness temperatures from automatic model-fitting procedure by fitting to the visibility data directly (Taylor et al.\ 2007).  
The minimum observable size for each source was calculated using equation (2) from Kovalev et al. (2005), where we computed the SNR of each core using the core flux density, the rms measured from the 5 GHz image, and a beam FWHM of 3 milliarcseconds (mas), the largest dimension of our restoring beam. 
For those sources where the estimated core size was less than this minimum size, we used the minimum size to compute the brightness temperature.

\begin{figure}
\plotone{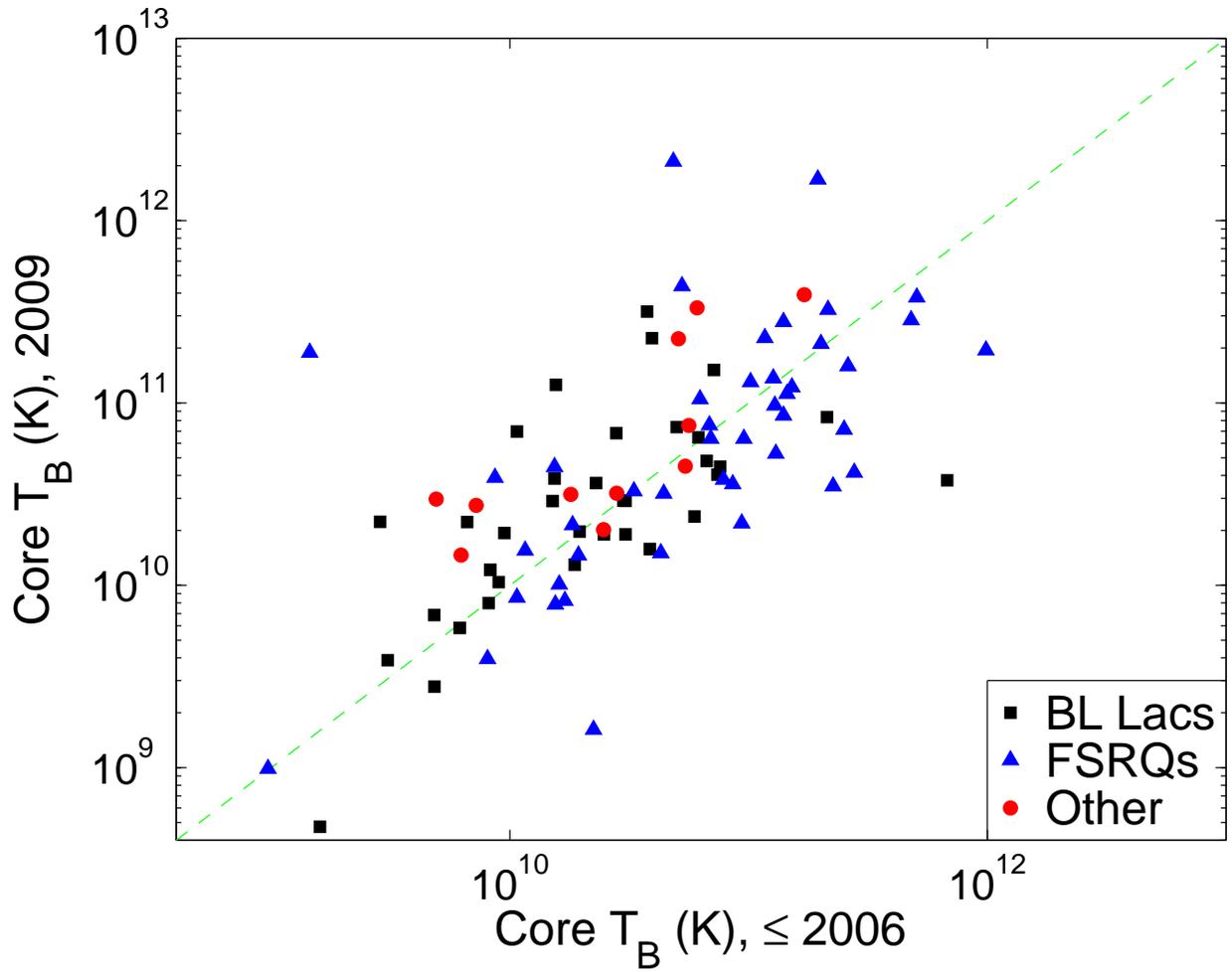}
\caption{Core brightness temperatures, current data (2009-2010) vs. archival data (prior to or during 2006).}
\label{ctbcomp}
\end{figure}

The core brightness temperatures are split almost evenly between those that are lower than they were prior to LAT detection and those that are higher.  Nearly all had core brightness temperatures in 2009 that were either within 5\% of the archival value or higher than the archival value.  The median core brightness temperature for the BL Lacs was 2.3$\times$10$^{10}$ K for the archival data and 2.9$\times$10$^{10}$ K for the new data.  The median core brightness temperature for the FSRQs in the archival data was 8.2$\times$10$^{10}$ K for and 6.4$\times$10$^{10}$ K for the new data.  However, the median absolute deviations were 6.34$\times$10$^{10}$ for the archival data and 5.13$\times$10$^{10}$ for the new data, so the difference in the medians is not a significant change.  See Fig.~\ref{ctbcomp} for a plot of the core flux densities in both epochs.

\subsection{Core Polarization Variability}

\begin{figure}
\plotone{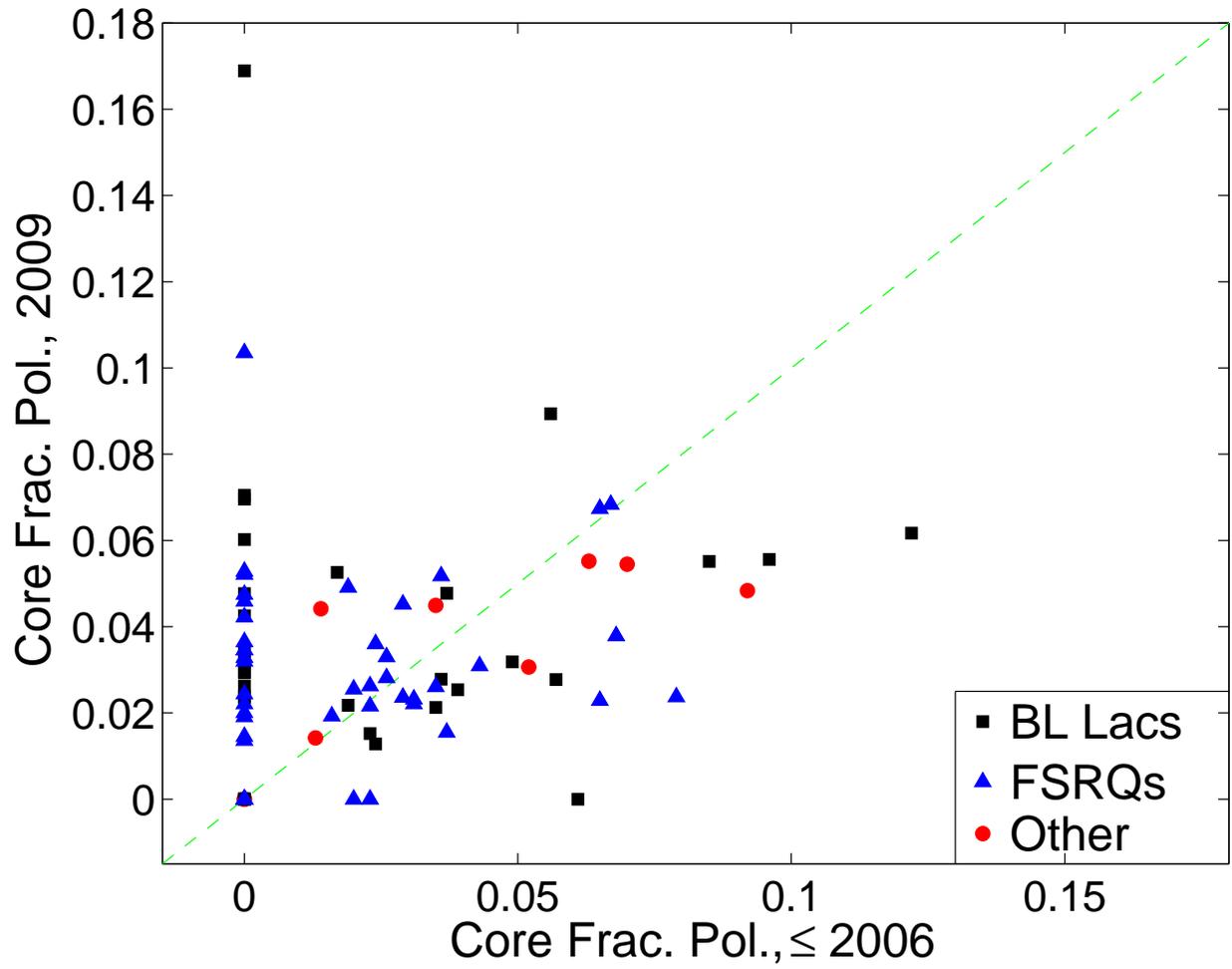}
\caption{Core fractional polarization at 5 GHz, current data (2009-2010) vs. archival data (prior to or during 2006).}
\label{cfpcomp}
\end{figure}

To measure the polarization properties of our sources, we used the Gaussian mask method described by Helmboldt et al. (2007).  In order to be considered a ``polarized'' source, the source had to have a polarized flux of at least 0.3\% of the Stokes I peak value (to avoid leakage contamination) and have at least a 5$\sigma$ detection (compared to the noise image generated by the AIPS task COMB).  We obtained measurements of core, jet, and total fractional polarization.  Of these, only the core fractional polarization yielded interesting results.  We detected polarization in the cores of 80\% (72 of 90) of the sources for which we had observations in two epochs.  Of our 90 sources, 48 showed higher core fractional polarization in the new data than the archival data.  This includes 17 BL Lac objects, 28 FSRQs, and 3 AGN/other objects.  Only 15 sources had no polarized flux in their cores in both epochs.  Twenty-seven of our sources showed a decrease in core fractional polarization in the new data including 11 BL Lac objects, 12 FSRQs, and 4 AGN/other objects.  Of the objects that showed a decrease in core fractional polarization, only 3 (1 BL Lac object and 2 FSRQs) were detected as polarized in the archival data and had no polarized flux in the core in the new data.
See Fig.~\ref{cfpcomp} for a plot showing how the core fractional polarizations have changed between the two epochs.

\section{LAT Detected versus non-LAT Detected}

Throughout this section we compare our large, contemporaneous sample with the non-LAT-detected sources in VIPS.

\subsection{Source Classes}

We used the automatic classification script from Helmboldt et al.\ (2007) to classify our sources by appearance for comparison with the VIPS sample.  We also went through them by eye and reclassified any objects that the script misidentified.  Sources were classified as point sources (PS), short jets (SJET), long jets (LJET), complex (CPLX), or compact symmetric object candidates (CSO).  LJET means the source has a jet with an angular extent of at least 6 mas.  SJET means the source has a discernible jet, but with an angular extent less than 6 mas.  PS means the source has no discernible jet.  CPLX indicates that the source has complicated structure beyond the typical point source or core-jet morphology.  See Table~\ref{morphtable} for a breakdown of the source classifications for both LAT and non-LAT sources.  From this table, it appears that $\gamma$-ray bright blazars are more likely to be LJET than the blazars not detected by the LAT.  Also, note the lack of CSO candidate objects among the $\gamma$-ray bright population.  This is something we also reported in Linford et al. (2011).

\subsection{Core Brightness Temperature}

\begin{figure}
\plotone{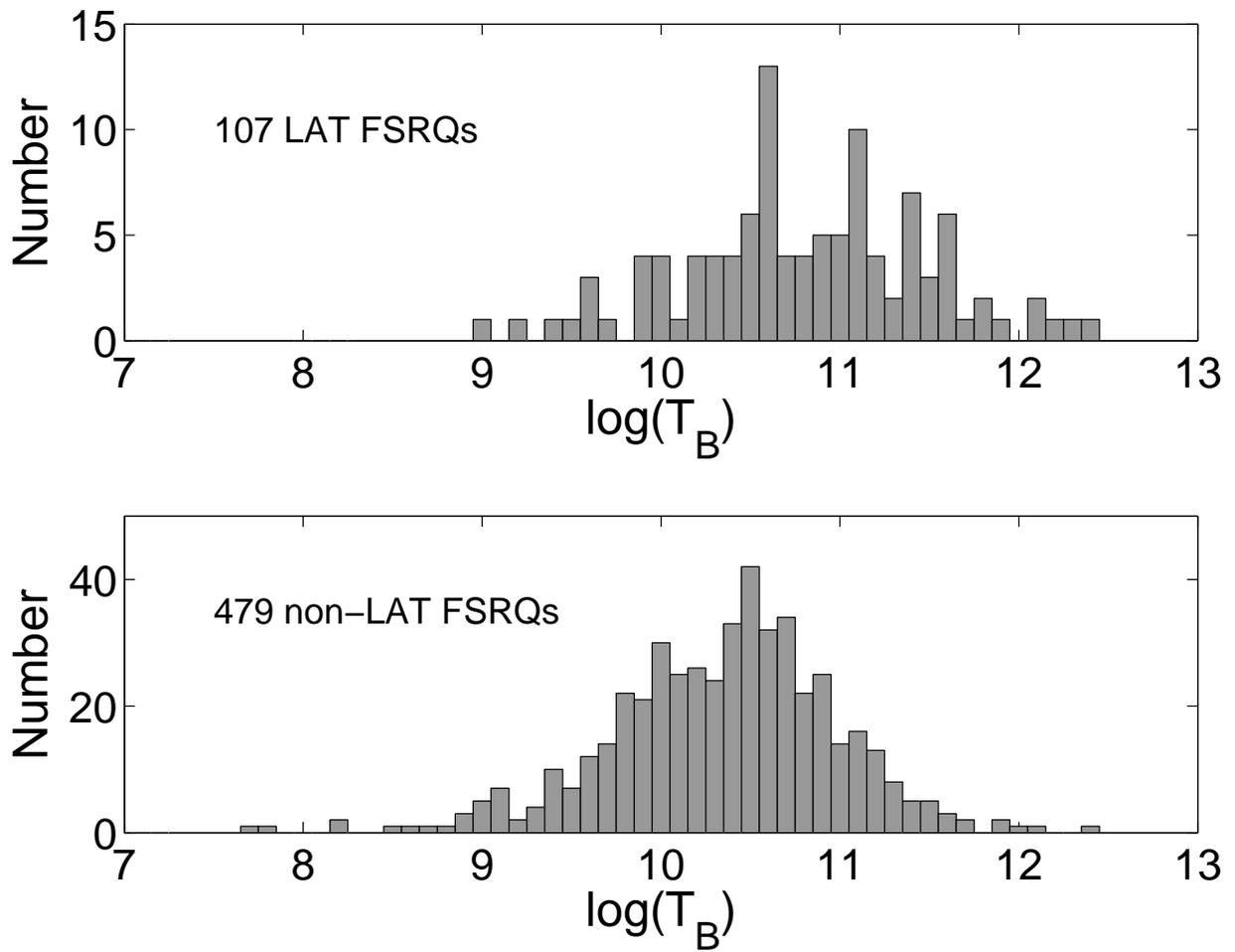}
\caption{
The distributions of core brightness temperatures for LAT-detected
FSRQs (top) compared to the distributions of core brightness temperature
for non-LAT FSRQs in VIPS (bottom).}
\label{FSRQTb}
\end{figure}

As discussed in Section 5.2, we obtained the core brightness temperatures by automatic model-fitting to the visibility data.  The LAT and non-LAT BL Lac objects were similar, with a median core brightness temperature of 2.89$\times$10$^{10}$ K for the LAT BL Lac objects and 2.65$\times$10$^{10}$ K for the non-LAT.  The FSRQs were quite different.  The median core brightness temperatures were 6.36$\times$10$^{10}$ K for the LAT FSRQs and 2.54$\times$10$^{10}$ K for the non-LAT FSRQs.  The K-S tests show that the LAT FSRQs are indeed very different from the non-LAT FSRQs, with only a 6$\times$10$^{-8}$ chance that the two distributions are drawn from the same parent population.  These results confirm those reported in Linford et al. (2011).  See Fig.~\ref{FSRQTb} for a plot of the core brightness temperature distributions of LAT and non-LAT FSRQs.

\subsection{Jet Brightness Temperature}

\begin{figure}
\plotone{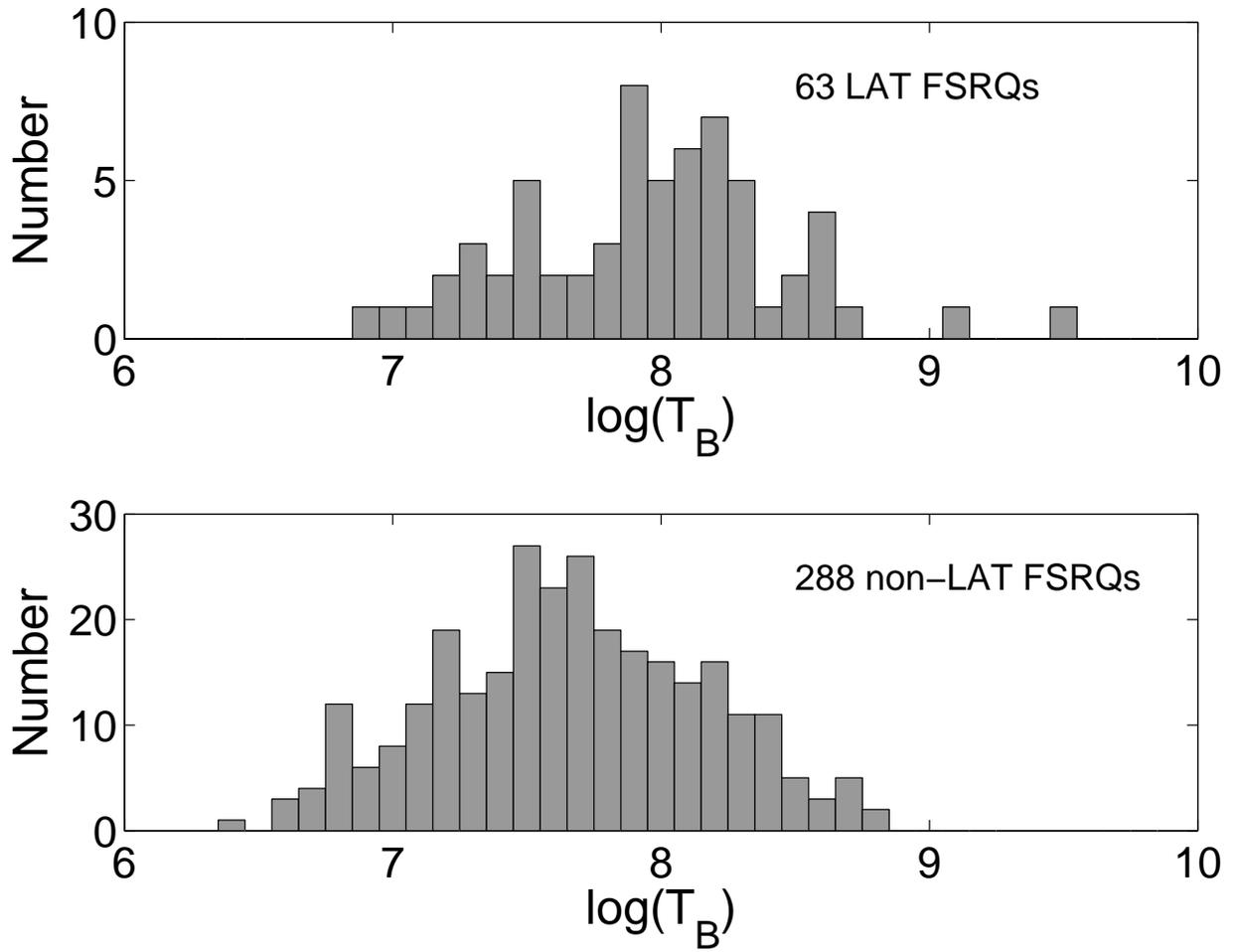}
\caption{The distributions for jet brightness temperatures for LAT (top) and non-LAT (bottom) FSRQs.}
\label{FSRQjt}
\end{figure}

Unlike the core brightness temperatures, the jet brightness temperatures (formally, the brightness temperature of the brightest jet component) were obtained by automatic model-fitting in the image data.  This was done because automatic visibility fitting to jet components has a tendency to go awry for complicated sources.  As with the core brightness temperatures, the LAT BL Lac objects were similar to the non-LAT BL Lac objects but the FSRQs appeared to be somewhat different.  The K-S test for the FSRQ jet brightness temperature distributions gave probability of 1$\times$10$^{-5}$ that the two were drawn from the same parent population.  See Fig.~\ref{FSRQjt} for the distributions of LAT and non-LAT FSRQ jet brightness temperatures.  The medians for the FSRQs with measured jet brightness temperatures were 9.8$\times$10$^{7}$ K for LAT sources and 4.5$\times$10$^{7}$ K for non-LAT sources.

\subsection{Jet Opening Angle, Bending, and Length}

We measured the mean apparent opening half angle for each source with core-jet morphology.  We used the following 
procedure (Taylor et al.\ 2007):  we measured the separation of each jet component from the core along the
jet axis (taken to be a linear fit to the component positions) and the
distance of each component from the jet axis, i.e., x' and y'
positions in a rotated coordinate system with the jet axis along the
x'-axis.  For each component, we measured its extent from its center
along a line perpendicular to the jet axis using the parameters of its
elliptical fit, and then deconvolved this using the extent of the
Gaussian restoring beam along the same line.  The opening half-angle
measured from each component is then taken to be 
\begin{equation}
\psi = {\rm arctan}[(|y'|+dr)/|x'|]
\end{equation}
 where $dr$ is the deconvolved Gaussian size
perpendicular to the jet axis.  After measuring this for each jet
component, we averaged them to get a single value.  This was only done
for sources with more than 2 total components, (i.e., at least 2 jet
components).  

\begin{figure}
\plotone{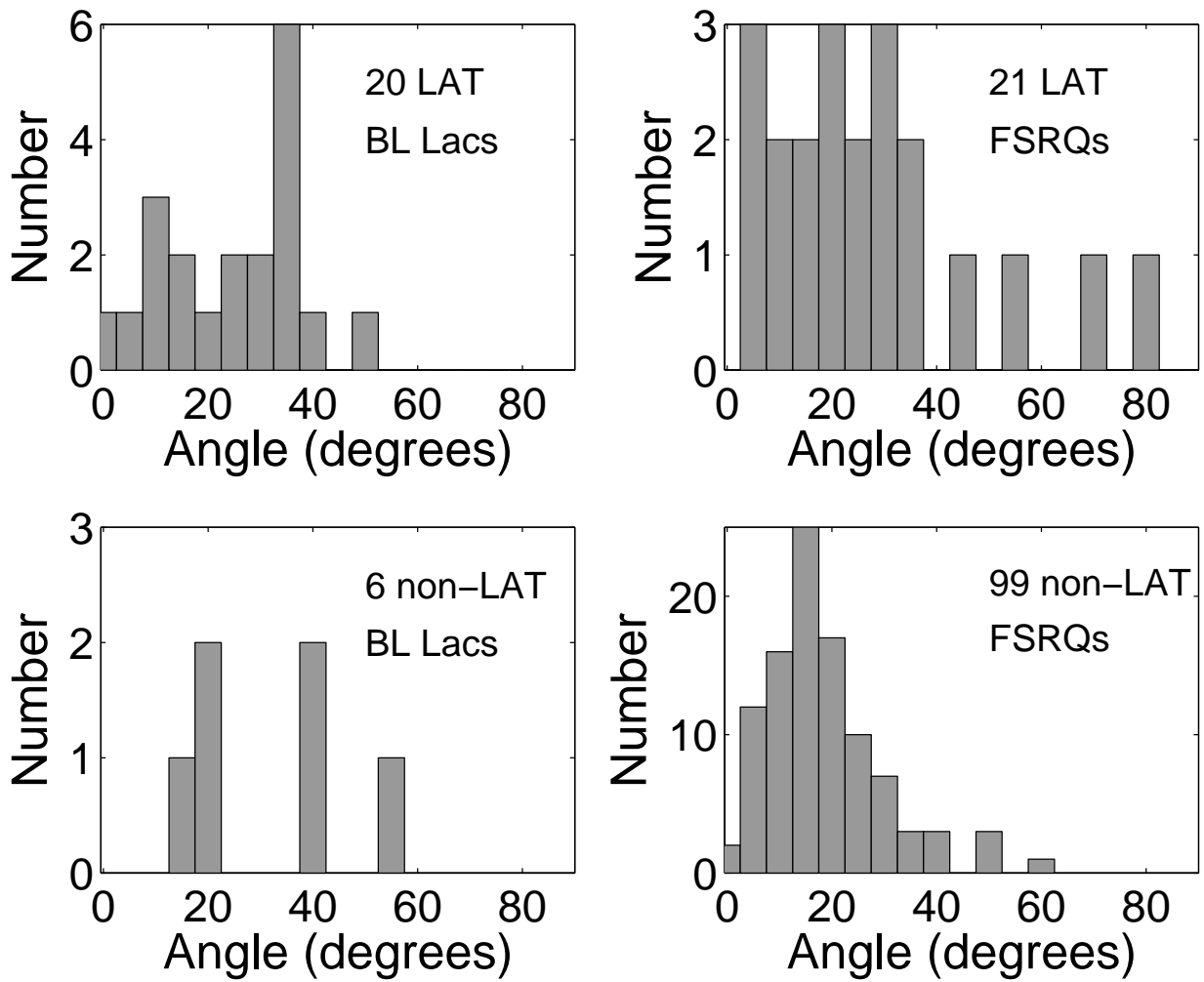}
\caption{The distributions of apparent opening angles for BL Lacs (left) and FSRQs (right) for both the LAT-detected (top) and non-LAT detected (bottom) sources.}
\label{comb_oa}
\end{figure}

The K-S tests for the LAT and non-LAT distributions found no significant differences between the two for any of the optical types.  However, we noticed that our LAT sources have large apparent opening angles ($\geq$ 30 degrees) more often than the non-LAT sources, 37\% of LAT sources versus 28\% of non-LAT sources.  See Fig.~\ref{comb_oa} for plots of the opening angle distributions.  Pushkarev et al. (2009) also reported that LAT-detected MOJAVE sources in the 3 month LAT catalog tended to have larger opening angles than non-LAT sources.  Ojha et al. (2010b) reported a tentative correlation between opening angle and $\gamma$-ray flux.  We did not find any strong evidence of this.  Using the Spearman test on all 49 of our sources with measured opening angles, we got a $\rho_{s}$ value of 0.2 and probability of getting the same $\rho_{s}$ value from random sampling of 16\%.  Looking at the FSRQs alone, we found a $\rho_{s}$ of 0.5 and a 1.4\% probability of getting the same result from random sampling.  Therefore, there may be a marginally significant correlation between $\gamma$-ray flux and opening angles for FSRQs.  Also, the K-S test result for the LAT and non-LAT FSRQs gave a 6.4\% chance that the two distributions are drawn from the same parent sample.  The BL Lacs showed no significant correlation and no significant difference between the LAT and non-LAT distributions.

In order to make a larger sample for the K-S test, we combined the BL Lac objects and FSRQs into one sample.  The result was a marginally significant difference between the LAT and non-LAT sources.  The K-S test result was a 0.4\% chance that the two distributions were drawn from the same parent population.  We also applied the Spearman test on this sample to look for correlation with LAT flux.  We did not find significant correlation, with a $\rho_{s}$ value of 0.3 and a probability of getting the same $\rho_{s}$ value from random sampling of 9\%.

Lister et al. (2011) reported a non-linear correlation between $\gamma$-ray loudness (the ratio of $\gamma$-ray luminosity to radio luminosity) and apparent opening angle.  They also found that all 19 of their sources with the large ($>$40$^{\circ}$) apparent opening angles had a $\gamma$-ray loudness of more than 100.  They did not find a significant difference in the opening angle distributions between BL Lac objects and FSRQs.  We are currently investigating these quantities in our data and we will report our results as soon as possible.

We also measure the change in jet position angle (``jet bending'', see Helmboldt et al. 2008) and the length of the jet (the maximum separation between the core component and the jet components).  We applied the K-S test to each of these measurements to see if the LAT and non-LAT distributions were different.  We did not find any significant differences for either of these properties.

\subsection{Polarization}

We measured the polarization characteristics of our sources the same way as described in Section 5.3.  As discussed therein, a source is considered to be polarized if it has a polarized flux of at least 0.3\% of the Stokes I peak and at least a 5$\sigma$ detection.  For the LAT sources, 176 of 232 (about 76\%) showed core polarization.  The FSRQs had the highest percentage of polarized sources with 96 of 107 (90\%).  The BL Lac objects had 67 of 95 (71\%) sources with core polarization, and the AGN/other sources had 13 of 30 (43\%) sources with core polarization.  Compare these numbers with the non-LAT VIPS sources where only 270 of 1018 (26.5\%) had core polarization.  For the non-LAT sample, the BL Lac objects were polarized most often, but only 10 of 24 (42\%).  The non-LAT FSRQs had 158 out of 479 (33\%) sources with core polarization, and the AGN/other only had 102 of 515 (20\%).

\begin{figure}
\plotone{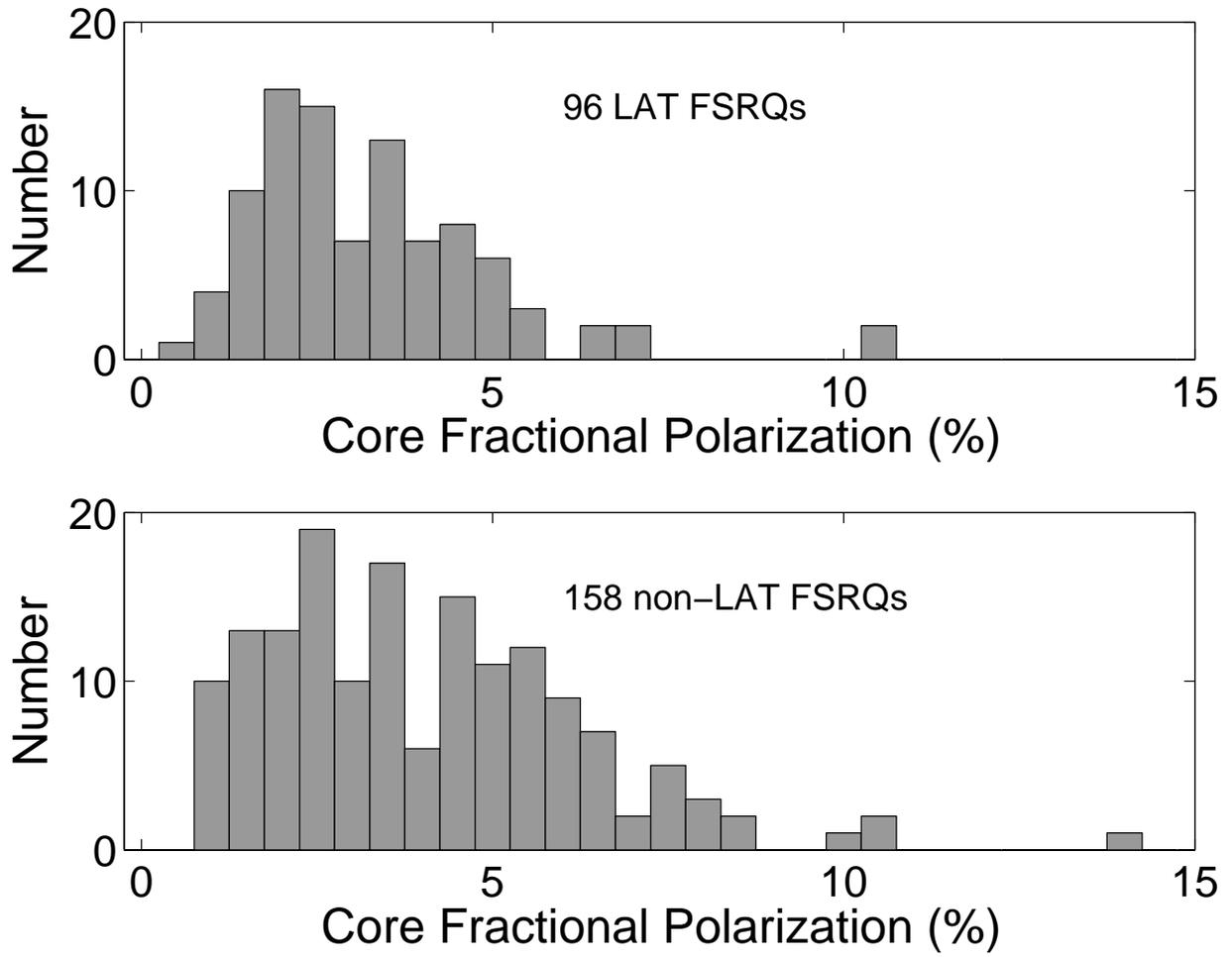}
\caption{The distribution of core fractional polarization for LAT (top) and non-LAT (bottom) FSRQs.}
\label{cfp_fsrq}
\end{figure}

Despite the LAT sources having polarized cores more often, they do not appear to be more strongly polarized.  The median core fractional polarization is 3.7\% for LAT BL Lac objects and 3.1\% for LAT FSRQs.  The non-LAT sample has higher medians, with 4.9\% for BL Lac objects and 3.7\% for FSRQs.  Overall, the median core fractional polarization for the LAT sample is 3.3\% while the median for the non-LAT sample is 4.4\%.  The K-S test showed no significant differences between the LAT and non-LAT BL Lac or AGN/other objects.  There is a marginally significant result for the FSRQs, with a probability of 0.2\% that the LAT and non-LAT FSRQs were drawn from the same parent population.  We plot the distribution of core fractional polarization for the FSRQs in Fig.~\ref{cfp_fsrq}.

\section{Notes on Individual Sources}
In this section we present information on low redshift (z$<$0.02) sources and non-blazar AGN in our sample.  Contour maps of these sources are shown in Fig.~\ref{imfig1}, except for F03197+4130 (a.k.a. 3C84) as it is a very well-known source.

\subsection{Low Redshift Objects}
Our sample contains 3 objects with redshifts of z$<$0.02.

\noindent
{\bf F03197+4130, 1FGL J0319.7+4130, class LJET:} This well known radio galaxy, 3C84, has a redshift of z = 0.018 (1LAC).  Its LAT flux was 213.59 $\pm$ 10.69 photons cm$^{-2}$ s$^{-1}$.  We measured a total VLBA flux density of 16.2 Jy.  Its core brightness temperature was 1.23$\times$10$^{11}$ K, which is relatively high for the AGN/Other objects.  We did not detect any polarized flux from this object.  This lack of strong polarization is well known and attributed to a Faraday screen consisting of the ionized gas which also produces the H$\alpha$ in the Perseus cluster (Taylor et al. 2006).  It has a jet and counterjet aligned north-south, with the northern jet showing significant free-free absorption (Walker et al. 2000).  We measured the southern jet to be about 15.8 mas long.  The northern counterjet had one dim component located about 11.8 mas from the core.  No contour map for this source is provided here as it is a very famous source.

\noindent
{\bf F09565+6938, 1FGL J0956.5+6938, class PS:} With a redshift of z=0.000677 (de Vaucouleurs et al. 1991), the starburst galaxy M82 is 1 of 2 starburst galaxies in the 1FGL.  It had a LAT flux of 38.40 $\pm$ 16.23 photons cm$^{-2}$ s$^{-1}$.  We found a total VLBA flux density of 14 mJy.  Its core brightness temperature was a relatively low 3.39$\times$10$^{9}$ K.  We did not detect any polarization for this object.  In the optical, this galaxy has a very striking x-shape with bright filaments running perpendicular through a spiral disk (e.g., Mutchler et al. 2007).  At 5 GHz, it is simply a point-source.  It is also thought to host two intermediate mass (12,000 to 43,000 $M_{\odot}$) black holes (Feng, Rao, \& Kaaret 2010).

\noindent
{\bf F17250+1151, 1FGL J1725.0+1151, class SJET:} This BL Lac object, also known as CRATES J1725+1152, has a redshift of z = 0.018 (Ciliegi, Bassani, \& Caroli 1993). Its LAT flux was 54.96 $\pm$ 23.25 photons cm$^{-2}$ s$^{-1}$.  We measured a total VLBA flux density of 63.3 mJy.  Its core fractional polarization was 3.5\%.  It is a compact source with a short jet extending about 3.5 mas to the southeast.  

\subsection{AGN}
The LAT 1FGL contains 28 sources which are classified as ``AGN'', which means ``other non-blazar AGN'' (Abdo et al. 2010a).  We have 12 of these objects in our sample. (Note: 1FGL J0319.7+4130, a.k.a. 3C 84, is discussed above as a low-redshift object.)  Some of the objects in this section are radio galaxies, but others may be misidentified FSRQs and BL Lac objects.

\noindent
{\bf J09235+4125, 1FGL J0923.2+4121, class LJET:} Also known as CRATES J0923+4125, this object had a LAT flux of 40.61 $\pm$ 10.01 photons cm$^{-2}$ s$^{-1}$.  Its redshift is z=0.028 (1LAC).  This object was observed in two epochs; first in May 2006 and second in November of 2009.  In 2006, we found a total VLBA flux density of 165 mJy.  In 2009, we found a total VLBA flux density of 220 mJy.    While its core brightness temperature is not unusually high or low for AGN/Other objects, it did change significantly between the two epochs.  In 2006, its core brightness temperature was 4.91$\times$10$^{9}$ K.  In 2009, we found a core brightness temperature of 2.97$\times$10$^{10}$ K.  We detected polarization in the core of this object in both epochs.  In 2006, its core fractional polarization was 5.2\%.  In 2009, its core fractional polarization was 3.1\%.  We also detected polarization in the jet in 2009, with a jet fractional polarization of 18.3\%.  For more discussion on this object, see Linford et al. (2011).  It has a jet extending about 9.7 mas to the east.  

\noindent
{\bf J12030+6031, 1FGL J1202.9+6032, class LJET:} Also known as CRATES J1203+6031, this object had a LAT flux of 44.79 $\pm$ 10.07 photons cm$^{-2}$ s$^{-1}$.  Its redshift is z=0.065 (1LAC).  This object was observed in two epochs; first in May of 2006 and second in December of 2009.  Its total VLBA flux density did not change much between the two epochs, but it did show an increase in core brightness temperature.  In 2006, its core brightness temperature was 7.22$\times$10$^{9}$.  In 2009, we found a core brightness temperature of 2.75$\times$10$^{10}$ K.  It had strong core polarization in both epochs.  In 2006, we found a core fractional polarization of 9.9\%.  In 2009, we found a core fractional polarization of 4.8\%. For more discussion on this object, see Linford et al. (2011).  It has a long jet extending about 10.5 mas to the south.  

\noindent
{\bf J13307+5202, 1FGL J1331.0+5202, class LJET:} This object is also called CRATES J1330+5202.  Its LAT flux was 41.52 photons cm$^{-2}$ s$^{-1}$, but is an upper limit.  It has a redshift of 0.688 (1LAC).  This object was observed in two epochs; first in July of 2006 and second in January of 2010. Its total VLBA flux density and core brightness temperature did not change significantly between the two epochs.  We did not detect any polarization from this object in either epoch. For further discussion on this object, see Linford et al. (2011).  It has a small jet extending about 7 mas to the southwest.  

\noindent
{\bf J16071+1551, 1FGL 1607.1+1552, class LJET:} Also called NVSS J160706+155134, its LAT flux was 33.90 $\pm$ 7.75 photons cm$^{-2}$ s$^{-1}$.    Its redshift is z=0.496 (1LAC).  It is called a BL Lac object in V\'eron-Cetty \& V\'eron (2006).  This object was observed in two epochs; the first in April 2006 and the second in January 2010.  In 2006, we measured a total VLBA flux density of 281 mJy.  In 2010, we measured a total VLBA flux density of 322 mJy.  It is a strongly polarized source in both epochs.  We measured a core fractional polarization of 3.5\% in 2006 and 4.5\% in 2010.  We also found polarization in the jet in both epochs.  In 2006 we measured a jet fractional polarization of 15.5\%, and in 2010 we measured 24\%.  For more information, see Linford et al. (2011).  It has a long jet extending about 10 mas to the east.  

\noindent
{\bf J16475+4950, 1FGL J1647.4+4948, class LJET:} Also called CRATES J1647+4950, this object had a LAT flux of 33.29 $\pm$ 8.54  photons cm$^{-2}$ s$^{-1}$.  It has a redshift of z=0.047 (1LAC).  This object was observed in two epochs; first in August of 2006 and second in January of 2010.  Its total VLBA flux density and core brightness temperature did not change significantly between those two epochs.  We did not detect any significant core polarization in this object in either epoch, but we did measure 27.1\% jet polarization in 2010.  For more information on this object, see Linford et al. (2011).  It has a small jet extending about 7 mas to the southeast.  

\noindent
{\bf J17240+4004, 1FGL J1724.0+4002, class LJET:} This object is also called CRATES J1724+4004. Its LAT flux was 47.16 $\pm$ 8.97 photons cm$^{-2}$ s$^{-1}$.  Its redshift is z=1.049 (1LAC).  It is tentatively classified as a BL Lac object in V\'eron-Cetty \& V\'eron (2006).  This object was observed in two epochs; first in February 1998 and second in January 2010.  Its total VLBA flux density did not change significantly between the two epochs, but its core brightness temperature more than doubled.  In 1998, its core brightness temperature was 6.08$\times$10$^{10}$.  In 2010, we found a core brightness temperature of 3.32$\times$10$^{11}$ K, which is the highest core brightness temperature for the ``AGN'' objects in our sample.  We did not detect any polarization in this object in either epoch.  For further discussion of this object, see Linford et al. (2011).  It has a long jet extending 9 mas to the northwest.  

\noindent
{\bf F02045+1516, 1FGL J0204.5+1516, class LJET:} This object is also known as 4C +15.05, and it is often classified as an FSRQ (V\'eron-Cetty \& V\'eron 2006, Savolainen et al. 2010, Agudo et al. 2010).  Its LAT flux was 29.98 $\pm$ 12.42 photons cm$^{-2}$ s$^{-1}$.  We measured a total VLBA flux density of 1.5 Jy.  Its redshift is z=0.405 (1LAC).  Its core brightness temperature was a relatively high 9.52$\times$10$^{10}$ K.  We did not detect any polarization from this object.  This is one of the sources monitored as part of the MOJAVE program.  Savolainen et al. (2010) reported an apparent velocity of 6.3c, a Doppler factor of 15.0, and a Lorentz factor of 8.9.  It has a bright jet component about 7 mas northwest of the core.  

\noindent
{\bf F03250+3403, 1FGL J0325.0+3402, class LJET:} Also called CRATES J0324+3410, this object is a Seyfert 1 galaxy (Abdo et al. 2009b).  It had a LAT flux of 56.29 $\pm$ 23.97 photons cm$^{-2}$ s$^{-1}$.  We measured a total VLBA flux density of 357 mJy.  It has a redshift of z=0.061 (1LAC).  We detected polarization in both the core and jet of this object.  Its core fractional polarization was 1.0\% and its jet fractional polarization was 33.4\%.  Zhou et al. (2007) reported the existence of spiral arm structure in this object based on \textit{Hubble Space Telescope} images.  Ant\'on et al. (2008), using the Nordic Optical Telescope, suggested that the apparent structure could be the result of a merger.  In the radio, it has a long straight jet extending about 9.6 mas to the southeast.  

\noindent
{\bf F16354+8228, 1FGL J1635.4+8228, class LJET:} This object is also known as NGC 6251.  It is an unabsorbed Seyfert 2, meaning it has a X-ray hydrogen column density $N_{H}\leq10^{22}$ cm$^{-2}$ and does not have a broad line region (Panessa \& Bassani 2002). It is a well-known radio source with an exceptionally long (about 200 kpc) and straight jet (Waggett, Warner, \& Baldwin 1977) and is popularly known as the ``blowtorch'' jet.  Its LAT flux was 45.51 $\pm$ 18.77 photons cm$^{-2}$ s$^{-1}$.  We measured a total VLBA flux density of 637 mJy.  Its redshift is z=0.025 (1LAC).  We did not detect any polarization in this object, although it has been reported to be weakly polarized (Chen et al. 2011).  It has a long straight jet extending about 8.5 mas to the northwest.  

\noindent
{\bf F16410+1143, 1FGL J1641.0+1143, LJET:} Also called NVSS J164058+114404, this object had a LAT flux of 47.36 $\pm$ 20.81 photons cm$^{-2}$ s$^{-1}$.  We measured a total VLBA flux density of 115 mJy.  It has a redshift of 0.078 (1LAC).  We found a core brightness temperature of 1.65$\times$10$^{9}$ K, which is the lowest of the ``AGN'' type objects.  We did not detect any polarization for this object. In the CRATES catalog, it is called a flat-spectrum radio source (Healey et al. 2007).  It has a jet extending about 10 mas to the west, ending in a semi-detached component.  

\noindent
{\bf F17566+5524, 1FGL J1756.6+5524, class LJET:} This object is also called CRATES J1757+5523.  It had a LAT flux of 27.79 $\pm$ 12.00 photons cm$^{-2}$ s$^{-1}$.  We measured a total VLBA flux density of 47 mJy.  Its redshift is 0.065 (1LAC). In the CRATES catalog, it is called a flat-spectrum radio source (Healey et al. 2007).  We did not detect any polarized flux from this object.  It has a jet stretching about 8 mas to the north-northwest.  

\subsection{AGN with No Optical ID}

Many sources in the LAT catalog are labeled as ``AGU'' sources, which means ``active galaxy of uncertain type'' (Abdo et al. 2010a).  These sources are associated with radio AGN, but have no optical identification as yet.  We have 16 of them in our sample.  It is likely that most of them are FSRQs.

\noindent
{\bf J11061+2812, 1FGL J1106.5+2809, class PS:} This object is also called CRATES J1106+2812.  The CRATES catalog lists it as a flat-spectrum radio source (Healey et al. 2007).  It had a LAT flux of 34.63 $\pm$ 15.03 photons cm$^{-2}$ s$^{-1}$.  It has a redshift of 0.847 (Adelman-McCarthy et al. 2008).  This object was observed in two epochs; first in February 2006 and second in December 2009.  In 2006, we found a total VLBA flux density of 276 mJy, and in 2009 we measured a total VLBA flux density of 227 mJy.  Its core brightness temperature more than doubled between the two epochs, going from 5.08$\times$10$^{10}$ K in 2006 to very high 2.25$\times$10$^{11}$ K in 2009.  In fact, this object had the highest core brightness temperature of all the ``AGU'' objects in our sample.  We detected polarization in the core of this object in both epochs.  In 2006, we measured a core fractional polarization of 1.4\%, and in 2009 its core fractional polarization was 4.4\%.  For more discussion on this object, see Linford et al. (2011).  It is a compact, point-source type object.  

\noindent
{\bf J11421+1547, 1FGL J1141.8+1549, class LJET:} This object is also called CRATES J1142+1547.  The CRATES catalog lists it as a flat-spectrum radio source (Healey et al. 2007).  It had a LAT flux of 12.87 $\pm$ 5.19 photons cm$^{-2}$ s$^{-1}$.  There is no published redshift for this object.  This object was observed in two epochs; first in February 2006 and second in December 2009.  Its total VLBA flux density was 172 mJy in 2006 and 139 mJy in 2009.  Its core brightness temperature increased between the two epochs from 5.62$\times$10$^{10}$ K to 9.15$\times$10$^{10}$ K, which is high for an AGN/Other type object.  We detected polarization in the core of this object in both epochs.  In 2006, the core fractional polarization was 6.3\% and in 2009 the core fractional polarization was 5.5\%.  We also detected polarization in the jet in 2009.  The jet fractional polarization was 31.1\%.  For more information on this object, see Linford et al. (2011).  It has a jet extending about 11.5 mas to the southeast.  

\noindent
{\bf J12248+4335 (class LJET) \& J12269+4340 (class SJET), 1FGL J1225.8+4336:} The LAT source 1FGL J1225.8+4336, with a LAT flux of 36.15 $\pm$ 15.51 photons cm$^{-2}$ s$^{-1}$, is associated with two VIPS sources, both of which were observed in two epochs; first in May 2006 and second in December 2009.  First, J12248+4335 (probability: 94\%, 1LAC), also called NVSS J122451+433520.  This object is called a BL Lac object by Plotkin et al. (2008) but listed as an ``Unknown'' type in both 1FGL and 1LAC.  It has a redshift of z=1.07491 (Sowards-Emmerd, Romani, \& Michelson 2003).  In 2006, we measured a total VLBA flux density of 174 mJy and in 2009 we measured a total VLBA flux density of 207 mJy.  Its core brightness temperature increased from 1.80$\times$10$^{10}$ K in 2006 to 3.15$\times$10$^{10}$ K in 2009.  We also detected polarization in the core and jet of this object in both epochs.  In 2006, its core fractional polarization was 7\% and its jet fractional polarization was 18\%.  In 2009, its core fractional polarization was 5.4\% and its jet fractional polarization was 42.6\%.  It has a jet extending about 8 mas to the northeast.
Second, J12269+4340 (probability: 87\%, 1LAC), or CRATES J1226+4340, is a FSRQ (V\'eron-Cetty \& V\'eron 2006 and 1LAC).  It has a redshift of z=2.002 (1LAC).  Its total VLBA flux density and core brightness temperature did not change significantly between the two epochs.  We detected core polarization only in the 2009 observation.  We found core fractional polarization of 4.2\%. It is a compact source with a possible jet extending about 5.6 mas to the south.  

\noindent
{\bf J13338+5057, 1FGL J1333.2+5056, class PS:} This object, also called CLASS J1333+5057, has no published classification.  Its LAT flux was 64.33 $\pm$ 27.63 photons cm$^{-2}$ s$^{-1}$.  It has a redshift of z = 1.362 (Abdo et al. 2009).  It was observed in two epochs; first in August 2006 and second in January 2010.  It was very dim in 2006, with a total VLBA flux density of 40 mJy, but it doubled to 80.2 mJy in 2010.  Its core brightness temperature also increased by about a factor of 2 from 6.24$\times$10$^{9}$ K in 2006 to 1.46$\times$10$^{10}$ K in 2010.  We did not detect any polarization in this object in either epoch.  It is a compact, point-source type object.  

\noindent
{\bf F03546+8009, 1FGL J0354.6+8009, class LJET:} This object is also known as CRATES J0354+8009.  The CRATES catalog lists it as a flat-spectrum radio source (Healey et al. 2007).  Its LAT flux was 46.05 $\pm$ 8.82 photons cm$^{-2}$ s$^{-1}$.  We measured a total VLBA flux density of 280 mJy.  There is no published redshift for this object.  It was a very strongly polarized source with a core fractional polarization of 14.2\%.  It has a broad, diffuse jet extending about 9.8 mas to the southeast.  It also had a large opening angle of 35$^{\circ}$.  Britzen et al. (2008) measured the kinematics of two of the jet components for this object and found apparent velocities of 0.007c for the one nearer the core and 0.064c or the one further from the core.  

\noindent
{\bf F05100+180A\&B (both class LJET), 1FGL J0510.0+1800:} The LAT source 1FGL J0510.0+1800, with a LAT flux of 33.38 $\pm$ 12.39 photons cm$^{-2}$ s$^{-1}$, is associated with 2 of our radio sources.  The first, F05100+180A, is also known as CRATES J0509+1806.  In 1LAC, its optical type is listed as ``Unknown''.  The CRATES catalog lists it as a flat-spectrum radio source (Healey et al. 2007).  It has no published redshift.  We measured a total VLBA flux density of 44.8 mJy.  We did not detect any polarized flux from this object.  It has a long jet extending to the east with a detached component about 16 mas from the center of the core.
The second radio source, F05100+180B, is a FSRQ also known as PKS 0446+17.  We measured a total VLBA flux density of 441 mJy.  We detected polarized flux in both the core and jet of this object.  Its core fractional polarization was 3.8\% and its jet fractional polarization was 12.8\%.  It has a long jet extending about 13 mas to the west.  

\noindent
{\bf F06544+5042, 1FGL J0654.4+5042, class SJET:} This object is also called CRATES J0654+5042.  The CRATES catalog lists it as a flat-spectrum radio source (Healey et al. 2007).  Its LAT flux was 42.02 $\pm$ 7.63 photons cm$^{-2}$ s$^{-1}$.  Abdo et al. (2010c) report that its $\gamma$-ray flux ($E>300$ MeV) is variable.  We measured a total VLBA flux density of 240 mJy.  It has no published redshift.  We found a core fractional polarization of 8.1\%.  It has a small jet extending about 4.7 mas to the east.  

\noindent
{\bf F08499+4852, 1FGL J0849.9+4852, class LJET:} This object is also known as CRATES J0850+4854.  The CRATES catalog lists it as a flat-spectrum radio source (Healey et al. 2007).  Its LAT flux was 29.87 $\pm$ 6.99 photons cm$^{-2}$ s$^{-1}$.  We measured a total VLBA flux density of 60.1 mJy.  There is no published redshift for this object.  We found a core brightness temperature of 1.20$\times$10$^{9}$ K, which is the lowest of all the AGN/Other type objects.  We did not detect any polarization in this object.  It has a small jet extending about 8 mas to the south.  It also has a large opening angle of 54.2$^{\circ}$.  

\noindent
{\bf F09055+1356, 1FGL J0905.5+1356, class SJET:} This object is also called CRATES J0905+1358.  The CRATES catalog lists it as a flat-spectrum radio source (Healey et al. 2007).  Its LAT flux was 18.39 $\pm$ 7.21 photons cm$^{-2}$ s$^{-1}$.  We measured a total VLBA flux density of 48.5 mJy.  It has no published redshift.  We did not detect any polarization in this object.  It is a compact object with a very short jet extending about 1.3 mas to the west.  

\noindent
{\bf F09498+1757, 1FGL J0949.8+1757, class SJET:} This object is also known as CRATES J0950+1804.  The CRATES catalog lists it as a flat-spectrum radio source (Healey et al. 2007).  It had an upper limit on its LAT flux of 18.85 photons cm$^{-2}$ s$^{-1}$.  We measured a total VLBA flux density of 37.2 mJy.  Its redshift is 0.69327 (Adelman-McCarthy et al. 2008).  We did not detect any polarized flux from this object.  It appears to have a short jet extending about 5 mas to the east. 

\noindent
{\bf F10485+7239, 1FGL J1048.5+7239, class LJET:} This object is also known as CRATES J1047+7238.  The CRATES catalog lists it as a flat-spectrum radio source (Healey et al. 2007).  Its LAT flux was 44.18 $\pm$ 19.09 photons cm$^{-2}$ s$^{-1}$.  We measured a total VLBA flux density of 61 mJy.  It has no published redshift.  We found a core fractional polarization of 5.2\%.  It has a jet extending about 12 mas to the west.  

\noindent
{\bf F13060+7852, 1FGL J1306.0+7852, class SJET:} This object is also called CRATES J1305+7854.  The CRATES catalog lists it as a flat-spectrum radio source (Healey et al. 2007).  Its LAT flux was 17.53 $\pm$ 6.82 photons cm$^{-2}$ s$^{-1}$.  We measured a total VLBA flux density of 213 mJy.  There is no published redshift for this object.  We found a core fractional polarization of 4.2\%.  It has a small jet extending about 5 mas to the northeast.  

\noindent
{\bf F13213+8310, 1FGL J1321.3+8310, class LJET:} This object is also called CRATES J1321+8316.  The CRATES catalog lists it as a flat-spectrum radio source (Healey et al. 2007).  Its LAT flux was 27.88 $\pm$ 11.62 photons cm$^{-2}$ s$^{-1}$.  We measured a total VLBA flux density of 232 mJy.  It has a redshift of z=1.024 (Britzen et al. 2007).  We did not detect any polarization in this object.  It has a long, diffuse jet extending about 22.5 mas to the west.  It also appears to have an older, detached jet component about 33 mas to the southwest.  Britzen et al. (2008) had measurements for the kinematics on two components in this source's jet, but for some reason they did not use the redshift quoted earlier.  Using that redshift, and Britzen's total proper motions of 0.041 mas yr$^{-1}$ for the component nearer to the core and 0.127 mas yr$^{-1}$ for the component further from the core, we find apparent velocities of 1.08c and 3.35c, respectively. 

\noindent
{\bf F19416+7214, 1FGL J1941.6+7214, class SJET:} This object is also called CRATES J1941+7221.  There is no published classification for it.  Its LAT flux was 66.46 $\pm$ 29.70 photons cm$^{-2}$ s$^{-1}$.  We measured a total VLBA flux density of 775 mJy.  There is no published redshift for this object.  We did not detect any polarization in this object.  It has a small jet extending about 4 mas to the south.  

\noindent
{\bf F20019+7040, 1FGL J2001.9+7040, class LJET:} This object is also known as CRATES J2001+7040.  The CRATES catalog lists it as a flat-spectrum radio source (Healey et al. 2007).  Its LAT flux was 27.18 $\pm$ 10.17 photons cm$^{-2}$ s$^{-1}$.  We measured a total VLBA flux density of 36.5 mJy.  There is no published redshift for this object.  We did not detect any polarization in this object.  It has a long, diffuse jet extending about 14 mas to the north.  

\noindent
{\bf F20497+1003, 1FGL J2049.7+1003, class LJET:} This object is also known as CRATES J2049+1003.  It may also be the EGRET source 3EG J2046+0933 (Bloom 2008).  The CRATES catalog lists it as a flat-spectrum radio source (Healey et al. 2007).  Its LAT flux was 58.69 $\pm$ 25.99 photons cm$^{-2}$ s$^{-1}$.  We measured a total VLBA flux density of 698 mJy.  It has no published redshift.  We found a core brightness temperature of 6.07$\times$10$^{10}$ K.  We found a very low core fractional polarization of 0.8\%.  It has a jet extending about 6.3 mas to the northwest. 

\subsection{Unidentified Sources}
Our sample contains 2 objects for which no optical classification is given in the 1FGL catalog.

\noindent
{\bf J09292+5013, 1FGL J0929.4+5000, class LJET:} In our previous paper (Linford et al. 2011), we identified this source as a BL Lac object.  It is associated with the BL Lac object VIPS J09292+5013 (CRATES J0929+5013), but with a low probability (67\%, 1LAC).  For this paper, we treated it as a BL Lac object.  It had a LAT flux of 21.40 $\pm$ 9.29 photons cm$^{-2}$ s$^{-1}$.  The two epochs for which we have observations of this object are February 1998 and November 2009.  It has a redshift of z=0.370387 (Adelman-McCarthy et al. 2005).  Its total VLBA flux density did not change significantly between the two epochs.  In 1998, we found a core brightness temperature of 6.67$\times$10$^{10}$ K, and in 2009 we found a core brightness temperature of 4.81$\times$10$^{10}$ K.  Even with this somewhat reduced core brightness temperature, it was still in the upper 33\% of BL Lac objects in the current sample.  It was a strongly polarized source with a core fractional polarization of 16.9\%. 

\noindent
{\bf J11540+6022, 1FGL J1152.1+6027, class PS:} This is associated with VIPS J11540+6022 (CRATES J1154+6022), a source of unknown type, with a probability of 73\% (1LAC). Healey et al. (2007) call it a flat-spectrum radio source.  Its LAT flux was 46.21 $\pm$ 20.76 photons cm$^{-2}$ s$^{-1}$.  We measured a total VLBA flux density of 232 mJy.  It has no published redshift.  We found a core brightness temperature of 3.93$\times$10$^{11}$ K, which is the highest core brightness temperature of all the ``Other'' type objects and the 16th highest in our entire sample.  We found a core fractional polarization of 1.4\%.

\section{Discussion and Conclusions}

\subsection{Polarization and Magnetic Fields}
The widely accepted picture of AGN central engine is a spinning super-massive black hole surrounded by an accretion disk (Blandford 1976, Lovelace 1976, Urry \& Padovani 1995).  In order to launch and collimate the jets, most models include a strong magnetic field which is coiled into a helical shape by the rotation of the accretion disk or the black hole.  Meier (2005) proposed a model where the magnetic field lines originate in the accretion disk and then thread through the ergosphere (the region near a rotating black hole where spacetime itself is rotating as a result of frame dragging) of the rotating central black hole.  This allows for the black hole to tightly wind the magnetic field lines, leading to reconnection events which can launch fast-moving material.  This may be the source of the $\gamma$-ray emitting regions in our blazars.

\begin{figure}
\plotone{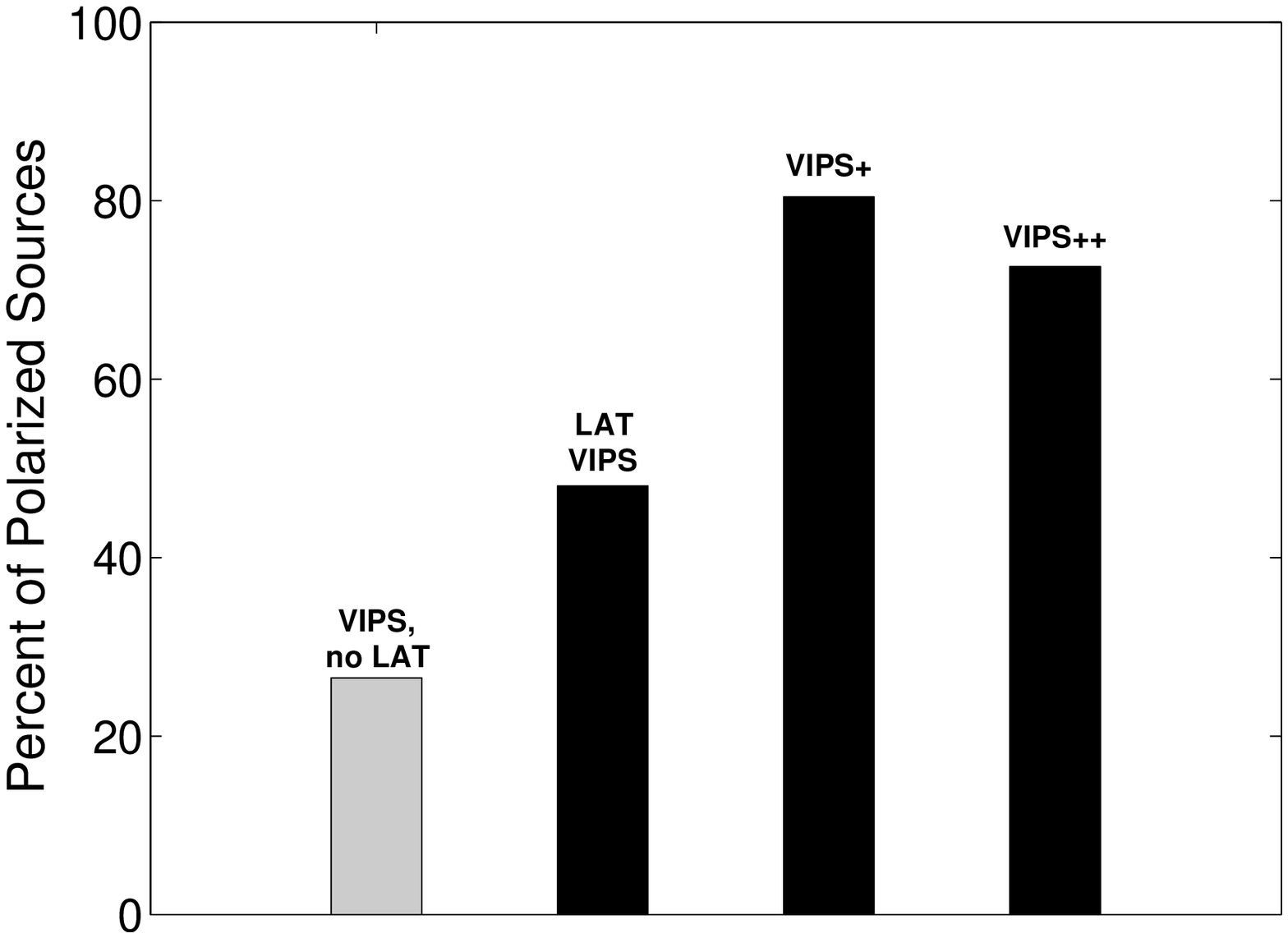}
\caption{The percentage of sources found to be polarized in various VLBI samples including VIPS (observations made prior to or during 2006),
VIPS+ (2nd epoch VIPS observations made contemporaneously with Fermi observations, 2009-2010), and VIPS++
(contemporaneous observations of additional LAT blazars, 2010). Gray indicates the sample with no LAT-detected sources, black indicates LAT samples.}
\label{polbar}
\end{figure}

We know that strong, uniform magnetic fields lead to polarized emission.  Because the majority (76\%) of our $\gamma$-ray bright sources showed significant polarization in their cores, it is obvious that they have strong, well-ordered magnetic fields in their centers.  The fact that the non-LAT sources are polarized less often (see Fig.~\ref{polbar}) leads us to believe that the strong, uniform magnetic fields are somehow tied to the $\gamma$-ray emission.  Also, recall from Section 5.3 that the core fractional polarization appeared to increase during LAT detection.  Furthermore, Abdo et al. (2010d) noted a dramatic change in the optical polarization orientation angle coincided with strong $\gamma$-ray emission in 3C279.  A possible explanation for this is that the magnetic fields in the cores become stronger (field lines wound more tightly) and more uniform (lines pulled into a less chaotic configuration) when $\gamma$-ray emission occurs.  In other words, following from the model in Meier 2005 discussed above, the spinning black hole winds up the magnetic fields lines until there is a reconnection event which launches a new, fast-moving jet component.  This new component upscatters background photons to GeV energies via inverse Compton processes (e.g., Bj\"ornsson 2010, Tavecchio et al. 2011, Abdo et al. 2011).  Meier (2005) expected several reconnection events in succession, followed by a time when the field has pulled back from the black hole.  Thus, the objects for which we do not detect polarization in the core could be in the state where the field is not being wound up by the black hole.

\subsection{Opening Angles}
While we did not have a large sample of objects with measured opening angles, we did find a hint that the LAT and non-LAT distributions are different.  The K-S test gave a 0.4\% probability that the LAT and non-LAT distributions of BL Lac object and FSRQ opening angles are taken from the same parent sample.  This is not a highly significant result, but it is tantalizing in view of what other studies have reported (e.g., Pushkarev et al. 2009; Ojha et al. 2010b; Lister et al. 2011).

\subsection{BL Lac Objects}

The only significant difference between the LAT and non-LAT BL Lac objects was that the LAT BL Lacs are polarized more often.  However, 10 of the 24 non-LAT BL Lac objects showed significant core polarization, so core polarization itself is not enough to separate the two populations.  It seems likely, therefore, that all BL Lacs produce $\gamma$-rays, but we simply do not detect all of them with the LAT.  It is well known that BL Lac objects are highly variable sources in radio and $\gamma$-ray bands (Abdo et al. 2011).  It is possible that LAT does not detect some BL Lac objects because they have lower than average Doppler factors, either as a result of lower velocities and/or jet orientations further from the line-of-sight.  Lister et al. (2009c) measured the kinematics of both LAT and non-LAT BL Lac objects, and found two non-LAT BL Lac objects with maximum jet material speeds higher than the LAT BL Lac objects.  However, they only had 21 BL Lac objects, 10 of which were LAT-detected.  Further study of BL Lac object kinematics is needed before we can claim LAT BL Lacs tend to have higher bulk material velocities.  It is also possible that BL Lac objects occasionally enter a state where $\gamma$-ray production ceases or is at least significantly reduced.  This question may be answered with continued monitoring of BL Lac objects with LAT and future instruments.

\subsection{FSRQs}

The LAT FSRQs exhibited several significant differences from the non-LAT FSRQs.  The LAT FSRQs had higher radio flux densities, higher core brightness temperatures, and were polarized much more often than the non-LAT FSRQs.  The LAT FSRQs also seemed to have higher jet brightness temperatures, although this is a marginal result.  It is also possible that the LAT FSRQs have larger opening angles than non-LAT FSRQs, but this result is tentative and is not strongly supported by the K-S test.

As with the BL Lac objects, we saw significant variation in flux density, core brightness temperature, and core polarization for those sources for which we had two epochs of observations.  However, as a group the FSRQs did not change significantly in flux density or core brightness temperature between the two epochs.  We did find an increase in the number of sources with strong core polarization.

About 90\% (96 of 107) of the LAT FSRQs showed significant polarization in their cores.  Compare this with only about 33\% of the non-LAT FSRQs showing core polarization, and we can see that core polarization is a strong indicator of $\gamma$-ray emission.  Of the 44 FSRQs for which we had observations in two epochs, 28 showed higher core fractional polarization during LAT detection.  Only 12 showed a decrease in core fractional polarization.  So, it appears that the cores of FSRQs tend to become more polarized during LAT detection.

\subsection{Other Non-Blazar AGN}

We found two significant differences between the LAT and non-LAT non-blazar AGN.  The first is the total radio flux density.  The LAT non-blazar AGN tended to have higher flux density than their non-LAT counterparts.  The K-S test resulted in a 8.8$\times$10$^{-4}$ chance that the LAT and non-LAT distributions are drawn from the same parent distribution.  Second, the LAT non-blazar AGN are polarized much more frequently.  We detected polarization in about 43\% (13 of 30) of our LAT sources, compared to only about 20\% (102 of 515) for our non-LAT sample.  As with the BL Lac objects and FSRQs, the non-blazar AGN are not necessarily more strongly polarized, but polarized more often.

Also, there remains a definite lack of CSO candidates among LAT-detected AGN, despite some predictions to the contrary (e.g., Stawarz et al. 2008).  

We thank the anonymous referee for helpful comments on the manuscript.  We thank Steve Tremblay and Marcello Giroletti for their helpful advice and comments.  We also thank Steve Myers and Josh Marvil for their help with obtaining EVLA observations of several sources for polarization angle calibration.
The National Radio Astronomy Observatory is a facility of the National Science Foundation operated under cooperative agreement by Associated Universities, Inc.  
This work made use of the Swinburne University of Technology software correlator, developed as part of the Australian Major National Research Facilities Programme and operated under licence.
The NASA/IPAC Extragalactic Database (NED) is operated by the Jet Propulsion Laboratory, California Institute of Technology, under contract with the National Aeronautics and Space Administration.  
We thank NASA for support under FERMI grant GSFC \#21078/FERMI08-0051 and the NRAO for support under Student Observing Support Award GSSP10-011.

\textwidth = 7.0truein
\textheight = 10.0truein
\begin{deluxetable}{cccrrccrrrrrrr}
\tablecolumns{14}
\tabletypesize{\scriptsize}
\tablewidth{0pt}
\rotate
\tablecaption{Source Data}
\tablehead{
\colhead{Source Name} & \colhead{1FGL Name} & \colhead{Alt. Name}	&	\colhead{RA}	&	\colhead{DEC}	&	\colhead{Opt.}	&	\colhead{Class}	&	\colhead{z}	&	\colhead{S$_5$}	&	\colhead{LAT} & \colhead{$\Delta$ LAT} & \colhead{Core T$_B$} & \colhead{Open.} & \colhead{$\Delta$ PA} \\
\colhead{} & \colhead{} & \colhead{} & \colhead{} &\colhead{} & \colhead{Type} & \colhead{} & \colhead{} & \colhead{} & \colhead{Flux} & \colhead{Flux} & \colhead{} & \colhead{Ang.} & \colhead{}}
\startdata
F00057+3815 & 	1FGL J0005.7+3815 & B2 0003+38A & 00:05:57.180 & 38:20:15.18 & 	bzq & 	LJET & 	0.229 & 	475 & 	38.61 & 	9.33 & 	3.59e+10 & 	\nodata & 	\nodata	\\
F00193+2017 & 	1FGL J0019.3+2017 & PKS 0017+200 & 00:19:37.850 & 20:21:45.61 & 	bzb & 	SJET & 	\nodata & 	680 & 	18.93 & 	7.63 & 	7.18e+10 & 	\nodata & 	\nodata	\\
F00230+4453 & 	1FGL J0023.0+4453 & B3 0020+446 & 00:23:35.441 & 44:56:35.81 & 	bzq & 	LJET & 	1.062 & 	120 & 	45.43 & 	20.21 & 	2.53e+10 & 	29 & 	9	\\
F00419+2318 & 	1FGL J0041.9+2318 & PKS 0039+230 & 00:42:04.550 & 23:20:01.21 & 	bzq & 	SJET & 	1.426 & 	676 & 	27.31 & 	11.19 & 	2.18e+10 & 	\nodata & 	\nodata	\\
F00580+3314 & 	1FGL J0058.0+3314 & CRATES J0058+3311 & 00:58:32.071 & 33:11:17.20 & 	bzb & 	PS & 	1.371 & 	151 & 	42.25 & 	17.77 & 	2.68e+11 & 	\nodata & 	\nodata	\\
F01022+4223 & 	1FGL J0102.2+4223 & CRATES J0102+4214 & 01:02:27.151 & 42:14:19.00 & 	bzq & 	PS & 	0.874 & 	163 & 	40.98 & 	9.25 & 	4.04e+11 & 	\nodata & 	\nodata	\\
F01090+1816 & 	1FGL J0109.0+1816 & CRATES J0109+1816 & 01:09:08.179 & 18:16:07.50 & 	bzb & 	LJET & 	0.145 & 	55.6 & 	17.20 & 	7.19 & 	1.62e+10 & 	\nodata & 	\nodata	\\
F01120+2247 & 	1FGL J0112.0+2247 & CGRaBS J0112+2244 & 01:12:05.820 & 22:44:38.80 & 	bzb & 	SJET & 	0.265 & 	327 & 	67.57 & 	7.96 & 	5.29e+10 & 	\nodata & 	\nodata	\\
F01129+3207 & 	1FGL J0112.9+3207 & 4C +31.03 & 01:12:50.330 & 32:08:17.59 & 	bzq & 	SJET & 	0.603 & 	405 & 	139.77 & 	9.04 & 	4.14e+10 & 	\nodata & 	\nodata	\\
F01138+4945 & 	1FGL J0113.8+4945 & CGRaBS J0113+4948 & 01:13:27.010 & 49:48:24.08 & 	bzq & 	LJET & 	0.389 & 	510 & 	37.30 & 	15.93 & 	3.49e+10 & 	7 & 	3	\\
F01144+1327 & 	1FGL J0114.4+1327\tablenotemark{*} & CRATES J0113+1324 & 01:13:54.511 & 13:24:52.49 & 	bzb & 	LJET & 	0.685 & 	106 & 	39.15 & 	7.82 & 	3.82e+9 & 	20 & 	1	\\
F01370+4751 & 	1FGL J0137.0+4751 & OC 457 & 01:36:58.591 & 47:51:29.09 & 	bzq & 	SJET & 	0.859 & 	3150 & 	193.96 & 	10.34 & 	2.61e+11 & 	\nodata & 	\nodata	\\
F01446+2703 & 	1FGL J0144.6+2703 & CRATES J0144+2705 & 01:44:33.559 & 27:05:03.08 & 	bzb & 	LJET & 	\nodata & 	305 & 	40.30 & 	8.50 & 	4.75e+10 & 	24 & 	-16	\\
F02035+7234 & 	1FGL J0203.5+7234 & CGRaBS J0203+7232 & 02:03:33.389 & 72:32:53.70 & 	bzb & 	LJET & 	\nodata & 	519 & 	63.95 & 	28.17 & 	8.35e+10 & 	\nodata & 	\nodata	\\
F02045+1516 & 	1FGL J0204.5+1516 & 4C +15.05 & 02:04:50.410 & 15:14:11.00 & 	agn & 	LJET & 	0.405 & 	1500 & 	29.98 & 	12.42 & 	9.52e+10 & 	\nodata & 	\nodata	\\
F02053+3217 & 	1FGL J0205.3+3217 & B2 0202+31 & 02:05:04.930 & 32:12:30.10 & 	bzq & 	LJET & 	1.466 & 	2120 & 	45.21 & 	19.21 & 	1.25e+12 & 	\nodata & 	\nodata	\\
F02112+1049 & 	1FGL J0211.2+1049 & CGRaBS J0211+1051 & 02:11:13.181 & 10:51:34.81 & 	bzb & 	SJET & 	\nodata & 	1060 & 	41.30 & 	8.45 & 	1.61e+11 & 	\nodata & 	\nodata	\\
F02178+7353 & 	1FGL J0217.8+7353 & 1ES 0212+735 & 02:17:30.821 & 73:49:32.59 & 	bzq & 	LJET & 	2.367 & 	3950 & 	77.11 & 	34.06 & 	1.28e+11 & 	\nodata & 	\nodata	\\
F02210+3555 & 	1FGL J0221.0+3555 & B2 0218+35 & 02:21:05.470 & 35:56:13.70 & 	bzq & 	CPLX & 	0.944 & 	835 & 	125.80 & 	9.39 & 	3.40e+9 & 	\nodata & 	\nodata	\\
F02308+4031 & 	1FGL J0230.8+4031 & B3 0227+403 & 02:30:45.710 & 40:32:53.09 & 	bzq & 	SJET & 	1.019 & 	501 & 	58.74 & 	26.55 & 	2.71e+10 & 	\nodata & 	\nodata	\\
F02379+2848 & 	1FGL J0237.9+2848 & 4C +28.07 & 02:37:52.411 & 28:48:09.00 & 	bzq & 	LJET & 	1.213 & 	2420 & 	135.56 & 	9.70 & 	1.50e+11 & 	27 & 	3	\\
F02386+1637 & 	1FGL J0238.6+1637 & PKS 0235+164 & 02:38:38.930 & 16:36:59.29 & 	bzb & 	PS & 	0.94 & 	919 & 	427.15 & 	12.63 & 	1.33e+11 & 	\nodata & 	\nodata	\\
F02435+7116 & 	1FGL J0243.5+7116 & CRATES J0243+7120 & 02:43:30.890 & 71:20:17.88 & 	bzb & 	SJET & 	\nodata & 	166 & 	25.10 & 	8.78 & 	9.43e+9 & 	\nodata & 	\nodata	\\
F02454+2413 & 	1FGL J0245.4+2413 & B2 0242+23 & 02:45:16.860 & 24:05:35.20 & 	bzq & 	LJET & 	2.243 & 	251 & 	41.30 & 	17.43 & 	2.75e+9 & 	\nodata & 	\nodata	\\
F02580+2033 & 	1FGL J0258.0+2033 & CRATES J0258+2030 & 02:58:07.310 & 20:30:01.58 & 	bzb & 	LJET & 	\nodata & 	66.3 & 	18.92 & 	7.78 & 	1.07e+10 & 	35 & 	-10	\\
F03106+3812 & 	1FGL J0310.6+3812 & B3 0307+380 & 03:10:49.879 & 38:14:53.81 & 	bzq & 	PS & 	0.816 & 	375 & 	41.19 & 	18.21 & 	2.57e+11 & 	\nodata & 	\nodata	\\
F03197+4130 & 	1FGL J0319.7+4130 & NGC 1275 & 03:19:48.161 & 41:30:42.12 & 	agn & 	LJET & 	0.018 & 	16200 & 	213.59 & 	10.69 & 	1.23e+11 & 	\nodata & 	\nodata	\\
F03250+3403 & 	1FGL J0325.0+3403 & B2 0321+33B & 03:24:41.160 & 34:10:45.80 & 	agn & 	LJET & 	0.061 & 	357 & 	56.29 & 	23.97 & 	2.15e+10 & 	\nodata & 	\nodata	\\
F03259+2219 & 	1FGL J0325.9+2219 & CGRaBS J0325+2224 & 03:25:36.809 & 22:24:00.40 & 	bzq & 	SJET & 	2.066 & 	889 & 	58.65 & 	24.82 & 	1.37e+11 & 	\nodata & 	\nodata	\\
F03546+8009 & 	1FGL J0354.6+8009 & CRATES J0354+8009 & 03:54:46.130 & 80:09:28.80 & 	agu & 	LJET & 	\nodata & 	280 & 	46.05 & 	8.82 & 	5.05e+10 & 	35 & 	-12	\\
F04335+2905 & 	1FGL J0433.5+2905 & CGRaBS J0433+2905 & 04:33:37.829 & 29:05:55.50 & 	bzb & 	SJET & 	\nodata & 	240 & 	58.78 & 	19.56 & 	4.17e+10 & 	\nodata & 	\nodata	\\
F04335+3230 & 	1FGL J0433.5+3230 & CRATES J0433+3237 & 04:33:40.690 & 32:37:12.00 & 	bzq & 	SJET & 	2.011 & 	59.8 & 	17.54 & 	7.08 & 	4.23e+9 & 	\nodata & 	\nodata	\\
F04406+2748 & 	1FGL J0440.6+2748 & B2 0437+27B & 04:40:50.369 & 27:50:46.79 & 	bzb & 	SJET & 	\nodata & 	31 & 	25.29 & 	0.00 & 	3.19e+9 & 	\nodata & 	\nodata	\\
F04486+112A &   1FGL J0448.6+1118\tablenotemark{*} & CRATES J0448+1127 & 04:48:50.410 & 11:27:28.58 &  bzq &   LJET &  1.369	&	383	&	60.13	&	23.11	&	1.23e+10	&	28	&	-36	\\
F04486+112B & 	1FGL J0448.6+1118\tablenotemark{*} & PKS 0446+11 & 04:49:07.670 & 11:21:28.58 & 	bzb & 	SJET & 	1.207\tablenotemark{N} & 	1260 & 	60.13 & 	23.11 & 	1.24e+11 & 	\nodata & 	\nodata	\\
F05092+1015 & 	1FGL J0509.2+1015 & PKS 0506+101 & 05:09:27.461 & 10:11:44.59 & 	bzq & 	LJET & 	0.621 & 	310 & 	53.24 & 	22.08 & 	4.00e+10 & 	\nodata & 	\nodata	\\
F05100+180A	&		1FGL J0510.0+1800\tablenotemark{*} & CRATES J0509+1806	&	05:09:42.910 & 18:06:30.31 &  agu &   LJET &  \nodata	&	44.8	&	32.38	&	12.39	&	8.80e+9	&	25	&	3	\\
F05100+180B & 	1FGL J0510.0+1800\tablenotemark{*} & PKS 0507+17 & 05:10:02.369 & 18:00:41.58 & 	bzq & 	LJET & 	0.416 & 	441 & 	32.38 & 	12.39 & 	3.63e+10 & 	22 & 	-3	\\
F05310+1331 & 	1FGL J0531.0+1331 & PKS 0528+134 & 05:30:56.419 & 13:31:55.09 & 	bzq & 	LJET & 	2.07 & 	2220 & 	146.45 & 	13.71 & 	1.34e+11 & 	\nodata & 	\nodata	\\
F06072+4739 & 	1FGL J0607.2+4739 & CGRaBS J0607+4739 & 06:07:23.249 & 47:39:47.02 & 	bzb & 	LJET & 	\nodata & 	226 & 	32.89 & 	13.16 & 	3.51e+10 & 	\nodata & 	\nodata	\\
F06127+4120 & 	1FGL J0612.7+4120 & B3 0609+413 & 06:12:51.190 & 41:22:37.42 & 	bzb & 	LJET & 	\nodata & 	296 & 	39.74 & 	13.60 & 	5.21e+10 & 	\nodata & 	\nodata	\\
F06169+5701 & 	1FGL J0616.9+5701 & CRATES J0617+5701 & 06:17:16.920 & 57:01:16.39 & 	bzb & 	LJET & 	\nodata & 	256 & 	16.24 & 	5.89 & 	8.23e+10 & 	\nodata & 	\nodata	\\
F06254+4440 & 	1FGL J0625.4+4440 & CGRaBS J0625+4440 & 06:25:18.259 & 44:40:01.60 & 	bzb & 	PS & 	\nodata & 	162 & 	21.27 & 	8.33 & 	2.25e+11 & 	\nodata & 	\nodata	\\
F06399+7325 & 	1FGL J0639.9+7325 & CGRaBS J0639+7324 & 06:39:21.960 & 73:24:58.00 & 	bzq & 	LJET & 	1.854 & 	612 & 	39.52 & 	17.33 & 	4.73e+10 & 	14 & 	4	\\
F06507+2503 & 	1FGL J0650.7+2503 & 1ES 0647+250 & 06:50:46.490 & 25:02:59.60 & 	bzb & 	PS & 	0.203 & 	21.5 & 	16.73 & 	5.54 & 	3.49e+9 & 	\nodata & 	\nodata	\\
F06544+5042 & 	1FGL J0654.4+5042 & CGRaBS J0654+5042 & 06:54:22.090 & 50:42:23.90 & 	agu & 	SJET & 	\nodata & 	240 & 	42.02 & 	7.63 & 	2.31e+10 & 	\nodata & 	\nodata	\\
F06543+4514 & 	1FGL J0654.3+4514 & B3 0650+453 & 06:54:23.710 & 45:14:23.50 & 	bzq & 	PS & 	0.933 & 	218 & 	115.24 & 	8.96 & 	1.67e+10 & 	\nodata & 	\nodata	\\
F07114+4731 & 	1FGL J0711.4+4731 & B3 0707+476 & 07:10:46.111 & 47:32:11.11 & 	bzb & 	LJET & 	1.292\tablenotemark{N} & 	620 & 	37.32 & 	15.00 & 	6.57e+10 & 	\nodata & 	\nodata	\\
F07127+5033 & 	1FGL J0712.7+5033 & CGRaBS J0712+5033 & 07:12:43.639 & 50:33:22.68 & 	bzb & 	PS & 	\nodata & 	252 & 	29.25 & 	10.21 & 	6.14e+10 & 	\nodata & 	\nodata	\\
F07193+3306 & 	1FGL J0719.3+3306 & B2 0716+33 & 07:19:19.421 & 33:07:09.70 & 	bzq & 	PS & 	0.779 & 	383 & 	70.34 & 	7.31 & 	9.91e+10 & 	\nodata & 	\nodata	\\
F07219+7120 & 	1FGL J0721.9+7120 & CGRaBS J0721+7120 & 07:21:53.450 & 71:20:36.38 & 	bzb & 	SJET & 	0.31 & 	1340 & 	149.23 & 	8.47 & 	7.42e+11 & 	\nodata & 	\nodata	\\
F07253+1431 & 	1FGL J0725.3+1431 & 4C +14.23 & 07:25:16.810 & 14:25:13.69 & 	bzq & 	LJET & 	1.038 & 	688 & 	37.66 & 	7.77 & 	3.80e+10 & 	37 & 	-54	\\
F07382+1741 & 	1FGL J0738.2+1741 & PKS 0735+178 & 07:38:07.390 & 17:42:19.01 & 	bzb & 	LJET & 	0.424 & 	644 & 	43.78 & 	13.96 & 	1.40e+10 & 	31 & 	10	\\
J07426+5444 & 	1FGL J0742.2+5443 & CRATES J0742+5444 & 07:42:39.7904 & 54:44:24.679 & 	bzq & 	PS & 	0.723 & 	220 & 	59.25 & 	24.76 & 	4.48e+11 & 	\nodata & 	\nodata	\\
J07464+2549 & 	1FGL J0746.6+2548 & B2 0743+25 & 07:46:25.8753 & 25:49:02.146 & 	bzq & 	SJET & 	2.979 & 	564 & 	42.28 & 	7.88 & 	3.79e+11 & 	\nodata & 	\nodata	\\
F07506+1235 & 	1FGL J0750.6+1235 & PKS 0748+126 & 07:50:52.051 & 12:31:04.80 & 	bzq & 	LJET & 	0.889 & 	3390 & 	37.46 & 	7.57 & 	3.30e+11 & 	\nodata & 	\nodata	\\
J07530+5352 & 	1FGL J0752.8+5353 & 4C +54.15 & 07:53:01.3847 & 53:52:59.636 & 	bzb & 	PS & 	0.2 & 	485 & 	27.01 & 	11.32 & 	3.76e+10 & 	\nodata & 	\nodata	\\
J08053+6144 & 	1FGL J0806.2+6148 & CGRaBS J0805+6144 & 08:05:18.1778 & 61:44:23.704 & 	bzq & 	SJET & 	3.033 & 	828 & 	36.04 & 	8.88 & 	5.28e+10 & 	\nodata & 	\nodata	\\
J08096+3455 & 	1FGL J0809.4+3455 & B2 0806+35 & 08:09:38.8868 & 34:55:37.248 & 	bzb & 	LJET & 	0.082 & 	58.6 & 	9.00 & 	3.34 & 	1.90e+10 & 	\nodata & 	\nodata	\\
J08098+5218 & 	1FGL J0809.5+5219 & CRATES J0809+5218 & 08:09:49.1899 & 52:18:58.252 & 	bzb & 	LJET & 	0.138 & 	108 & 	43.03 & 	7.77 & 	2.78e+9 & 	50 & 	20	\\
J08146+6431 & 	1FGL J0815.0+6434 & CGRaBS J0814+6431 & 08:14:39.1912 & 64:31:22.04 & 	bzb & 	LJET & 	\nodata & 	102 & 	32.76 & 	12.91 & 	7.31e+10 & 	\nodata & 	\nodata	\\
J08163+5739 & 	1FGL J0816.7+5739 & BZB J0816+5739 & 08:16:23.8223 & 57:39:09.509 & 	bzb & 	CPLX & 	\nodata & 	21.5 & 	15.19 & 	5.65 & 	4.77e+8 & 	\nodata & 	\nodata	\\
J08182+4222 & 	1FGL J0818.2+4222 & B3 0814+425 & 08:18:15.9995 & 42:22:45.408 & 	bzb & 	SJET & 	0.53\tablenotemark{N} & 	1660 & 	108.47 & 	8.38 & 	3.17e+11 & 	\nodata & 	\nodata	\\
J08247+5552 & 	1FGL J0825.0+5555 & OJ 535 & 08:24:47.2364 & 55:52:42.662 & 	bzq & 	LJET & 	1.417 & 	971 & 	62.43 & 	8.92 & 	3.80e+10 & 	17 & 	-11	\\
J08308+2410\tablenotemark{M} & 	1FGL J0830.5+2407 & OJ 248 & 08:30:52.0855 & 24:10:59.818 & 	bzq & 	SJET & 	0.94 & 	1110 & 	65.29 & 	8.21 & 	4.14e+11 & 	\nodata & 	\nodata	\\
J08338+4224 & 	1FGL J0834.4+4221 & B3 0830+425 & 08:33:53.8852 & 42:24:01.859 & 	bzq & 	SJET & 	0.249 & 	249 & 	31.44 & 	7.80 & 	3.90e+10 & 	\nodata & 	\nodata	\\
F08422+7054 & 	1FGL J0842.2+7054 & 4C +71.07 & 08:41:24.360 & 70:53:42.22 & 	bzq & 	LJET & 	2.218 & 	1680 & 	61.94 & 	8.92 & 	9.87e+9 & 	9 & 	-19	\\
F08499+4852 & 	1FGL J0849.9+4852 & CRATES J0850+4854 & 08:50:00.530 & 48:54:52.60 & 	agu & 	LJET & 	\nodata & 	60.1 & 	29.87 & 	6.99 & 	1.20e+9 & 	54 & 	121	\\
J08548+2006\tablenotemark{M} & 	1FGL J0854.8+2006 & OJ 287 & 08:54:48.8741 & 20:06:30.639 & 	bzb & 	SJET & 	0.306 & 	2380 & 	63.31 & 	13.83 & 	3.17e+11 & 	\nodata & 	\nodata	\\
J08566+2057 & 	1FGL J0856.6+2103\tablenotemark{*} & CRATES J0850+2057 & 08:56:39.7398 & 20:57:43.426 & 	bzq & 	LJET & 	0.539\tablenotemark{N} & 	45.1 & 	39.89 & 	17.57 & 	1.28e+10 & 	16 & 	10	\\
J08569+2111 & 	1FGL J0856.6+2103\tablenotemark{*} & OJ 290 & 08:56:57.2424 & 21:11:43.64 & 	bzq & 	LJET & 	2.098 & 	365 & 	39.89 & 	17.57 & 	8.24e+9 & 	\nodata & 	\nodata	\\
F09055+1356 & 	1FGL J0905.5+1356 & CRATES J0905+1358 & 09:05:34.990 & 13:58:06.31 & 	agu & 	SJET & 	\nodata & 	48.5 & 	18.39 & 	7.21 & 	7.33e+9 & 	\nodata & 	\nodata	\\
J09106+3329 & 	1FGL J0910.7+3332 & Ton 1015 & 09:10:37.0354 & 33:29:24.418 & 	bzb & 	LJET & 	0.354 & 	130 & 	22.76 & 	9.04 & 	4.04e+10 & 	\nodata & 	\nodata	\\
J09121+4126 & 	1FGL J0912.3+4127 & B3 0908+416B & 09:12:11.6174 & 41:26:09.356 & 	bzq & 	LJET & 	2.563 & 	132 & 	18.51 & 	7.61 & 	3.95e+9 & 	\nodata & 	\nodata	\\
J09158+2933 & 	1FGL J0915.7+2931 & B2 0912+29 & 09:15:52.4014 & 29:33:23.982 & 	bzb & 	LJET & 	\nodata & 	80 & 	22.64 & 	8.34 & 	1.30e+10 & 	\nodata & 	\nodata	\\
J09209+4441 & 	1FGL J0920.9+4441 & B3 0917+449 & 09:20:58.4599 & 44:41:53.988 & 	bzq & 	LJET & 	2.19 & 	1020 & 	242.01 & 	9.44 & 	3.59e+10 & 	44 & 	-159	\\
J09216+6215 & 	1FGL J0919.6+6216 & OK 630 & 09:21:36.2322 & 62:15:52.185 & 	bzq & 	LJET & 	1.446 & 	1410 & 	35.59 & 	7.95 & 	1.59e+11 & 	\nodata & 	\nodata	\\
J09235+4125 & 	1FGL J0923.2+4121 & B3 0920+416 & 09:23:31.3037 & 41:25:27.429 & 	agn & 	LJET & 	0.028 & 	220 & 	40.61 & 	10.01 & 	2.97e+10 & 	10 & 	1	\\
J09238+2815 & 	1FGL J0924.2+2812 & B2 0920+28 & 09:23:51.522 & 28:15:24.966 & 	bzq & 	PS & 	0.744 & 	899 & 	44.39 & 	19.61 & 	1.95e+11 & 	\nodata & 	\nodata	\\
J09292+5013 & 	1FGL J0929.4+5000 & CRATES J0929+5013 & 09:29:15.4401 & 50:13:35.982 & 	bzb & 	LJET & 	0.37039\tablenotemark{N} & 	430 & 	21.40 & 	9.29 & 	4.81e+10 & 	\nodata & 	\nodata	\\
J09341+3926 & 	1FGL J0934.5+3929 & CGRaBS J0934+3926 & 09:34:06.674 & 39:26:32.125 & 	bzb & 	PS & 	0.044\tablenotemark{N} & 	108 & 	19.72 & 	7.73 & 	2.89e+10 & 	\nodata & 	\nodata	\\
J09372+5008 & 	1FGL J0937.7+5005 & CGRaBS J0937+5008 & 09:37:12.3257 & 50:08:52.082 & 	bzq & 	PS & 	0.276 & 	324 & 	52.33 & 	11.77 & 	1.05e+11 & 	\nodata & 	\nodata	\\
J09418+2728 & 	1FGL J0941.2+2722 & CGRaBS J0941+2728 & 09:41:48.1135 & 27:28:38.818 & 	bzq & 	PS & 	1.306 & 	235 & 	18.15 & 	7.24 & 	1.46e+10 & 	\nodata & 	\nodata	\\
F09456+5754 & 	1FGL J0945.6+5754 & CRATES J0945+5757 & 09:45:42.240 & 57:57:47.70 & 	bzb & 	LJET & 	0.229 & 	42.9 & 	17.96 & 	7.22 & 	1.64e+10 & 	\nodata & 	\nodata	\\
F09466+1012 & 	1FGL J0946.6+1012 & CRATES J0946+1017 & 09:46:35.071 & 10:17:06.11 & 	bzq & 	SJET & 	1.007 & 	245 & 	28.22 & 	11.71 & 	2.42e+10 & 	\nodata & 	\nodata	\\
J09496+1752 & 	1FGL J0949.8+1757\tablenotemark{*} & CRATES J0949+1752 & 09:49:39.7634 & 17:52:49.432 & 	bzq & 	LJET & 	0.693 & 	207 & 	18.85 & 	0.00 & 	1.50e+11 & 	\nodata & 	\nodata	\\
F09498+1757 & 	1FGL J0949.8+1757\tablenotemark{*} & CRATES J0950+1804 & 09:50:00.310 & 18:04:18.70 & 	agu & 	SJET & 	0.69327\tablenotemark{N} & 	37.2 & 	18.85 & 	0.00 & 	1.43e+9 & 	\nodata & 	\nodata	\\
F09565+6938 & 	1FGL J0956.5+6938 & M 82 & 09:55:52.726 & 69:40:45.77 & 	sbg & 	PS & 	0.000677\tablenotemark{N} & 	14 & 	38.40 & 	16.23 & 	3.39e+9 & 	\nodata & 	\nodata	\\
J09568+2515 & 	1FGL J0956.9+2513 & B2 0954+25A & 09:56:49.8747 & 25:15:16.047 & 	bzq & 	PS & 	0.712 & 	645 & 	28.07 & 	11.59 & 	7.88e+9 & 	\nodata & 	\nodata	\\
J09576+5522 & 	1FGL J0957.7+5523 & 4C +55.17 & 09:57:38.1837 & 55:22:57.74 & 	bzq & 	LJET & 	0.896 & 	580 & 	106.45 & 	7.43 & 	9.88e+8 & 	68 & 	-1	\\
F10001+6539 & 	1FGL J1000.1+6539 & CGRaBS J0958+6533 & 09:58:47.251 & 65:33:54.79 & 	bzb & 	LJET & 	0.367 & 	1450 & 	28.21 & 	11.72 & 	5.07e+11 & 	\nodata & 	\nodata	\\
F10127+2440 & 	1FGL J1012.7+2440 & CRATES J1012+2439 & 10:12:41.381 & 24:39:23.40 & 	bzq & 	SJET & 	1.805 & 	63.9 & 	49.66 & 	6.81 & 	5.56e+9 & 	\nodata & 	\nodata	\\
J10150+4926 & 	1FGL J1015.1+4927 & 1ES 1011+496 & 10:15:04.1336 & 49:26:00.704 & 	bzb & 	LJET & 	0.2 & 	201 & 	63.31 & 	6.16 & 	7.98e+9 & 	\nodata & 	\nodata	\\
J10330+4116 & 	1FGL J1033.2+4116 & B3 1030+415 & 10:33:03.7086 & 41:16:06.234 & 	bzq & 	LJET & 	1.117 & 	1290 & 	29.30 & 	6.86 & 	2.22e+12 & 	\nodata & 	\nodata	\\
J10338+6051 & 	1FGL J1033.8+6048 & CGRaBS J1033+6051 & 10:33:51.427 & 60:51:07.342 & 	bzq & 	SJET & 	1.401 & 	188 & 	50.15 & 	7.02 & 	1.55e+10 & 	\nodata & 	\nodata	\\
F10377+5711 & 	1FGL J1037.7+5711 & CRATES J1037+5711 & 10:37:44.311 & 57:11:55.61 & 	bzb & 	PS & 	\nodata & 	61.8 & 	30.14 & 	10.02 & 	1.73e+10 & 	\nodata & 	\nodata	\\
J10431+2408 & 	1FGL J1043.1+2404 & B2 1040+24A & 10:43:09.0347 & 24:08:35.43 & 	bzb & 	SJET & 	0.56 & 	738 & 	18.02 & 	6.92 & 	1.26e+11 & 	\nodata & 	\nodata	\\
F10487+8054 & 	1FGL J1048.7+8054 & CGRaBS J1044+8054 & 10:44:23.071 & 80:54:39.38 & 	bzq & 	LJET & 	1.26 & 	727 & 	57.94 & 	8.73 & 	3.32e+10 & 	3 & 	-10	\\
F10485+7239 & 	1FGL J1048.5+7239 & CRATES J1047+7238 & 10:47:47.520 & 72:38:13.02 & 	agu & 	LJET & 	\nodata & 	61 & 	44.18 & 	19.09 & 	1.60e+10 & 	\nodata & 	\nodata	\\
F10488+7145 & 	1FGL J1048.8+7145 & CGRaBS J1048+7143 & 10:48:27.619 & 71:43:35.90 & 	bzq & 	PS & 	1.15 & 	1170 & 	59.88 & 	26.71 & 	1.09e+11 & 	\nodata & 	\nodata	\\
J10586+5628 & 	1FGL J1058.6+5628 & CGRaBS J1058+5628 & 10:58:37.7261 & 56:28:11.18 & 	bzb & 	LJET & 	0.143 & 	159 & 	56.29 & 	6.84 & 	1.93e+10 & 	\nodata & 	\nodata	\\
J11044+3812 & 	1FGL J1104.4+3812 & Mkn 421 & 11:04:27.3145 & 38:12:31.794 & 	bzb & 	LJET & 	0.03 & 	262 & 	170.63 & 	7.25 & 	8.35e+10 & 	\nodata & 	\nodata	\\
J11061+2812 & 	1FGL J1106.5+2809 & CRATES J1106+2812 & 11:06:07.2592 & 28:12:47.045 & 	agu & 	PS & 	0.847\tablenotemark{N} & 	227 & 	34.63 & 	15.03 & 	2.25e+11 & 	\nodata & 	\nodata	\\
J11126+3446 & 	1FGL J1112.8+3444 & CRATES J1112+3446 & 11:12:38.7673 & 34:46:39.124 & 	bzq & 	LJET & 	1.949 & 	197 & 	33.70 & 	13.80 & 	6.37e+10 & 	\nodata & 	\nodata	\\
F11171+2013 & 	1FGL J1117.1+2013 & CRATES J1117+2014 & 11:17:06.259 & 20:14:07.40 & 	bzb & 	LJET & 	0.138 & 	36.5 & 	11.66 & 	3.57 & 	1.28e+10 & 	\nodata & 	\nodata	\\
J11240+2336 & 	1FGL J1123.9+2339 & OM 235 & 11:24:02.7109 & 23:36:45.876 & 	bzb & 	SJET & 	\nodata & 	362 & 	23.95 & 	9.65 & 	6.96e+10 & 	\nodata & 	\nodata	\\
F11366+7009 & 	1FGL J1136.6+7009 & Mkn 180 & 11:36:26.410 & 70:09:27.32 & 	bzb & 	LJET & 	0.045 & 	136 & 	13.12 & 	4.46 & 	2.51e+10 & 	\nodata & 	\nodata	\\
J11421+1547 & 	1FGL J1141.8+1549 & CRATES J1142+1547 & 11:42:07.7378 & 15:47:54.202 & 	agu & 	LJET & 	\nodata & 	139 & 	12.87 & 	5.19 & 	9.15e+10 & 	28 & 	5	\\
J11469+3958 & 	1FGL J1146.8+4004 & B2 1144+40 & 11:46:58.2987 & 39:58:34.307 & 	bzq & 	PS & 	1.089 & 	603 & 	46.91 & 	20.17 & 	2.85e+11 & 	\nodata & 	\nodata	\\
J11503+2417 & 	1FGL J1150.2+2419 & B2 1147+24 & 11:50:19.2146 & 24:17:53.852 & 	bzb & 	LJET & 	0.2 & 	621 & 	29.11 & 	11.81 & 	4.47e+10 & 	20 & 	-8	\\
J11514+5859 & 	1FGL J1151.6+5857 & CRATES J1151+5859 & 11:51:24.6554 & 58:59:17.552 & 	bzb & 	SJET & 	\nodata & 	61.7 & 	18.22 & 	7.12 & 	3.88e+9 & 	\nodata & 	\nodata	\\
J11540+6022 & 	1FGL J1152.1+6027 & CRATES J1154+6022 & 11:54:04.5339 & 60:22:20.785 & 	\nodata & 	PS & 	\nodata & 	232 & 	46.21 & 	20.76 & 	3.93e+11 & 	\nodata & 	\nodata	\\
J11595+2914\tablenotemark{M} & 	1FGL J1159.4+2914 & 4C +29.45 & 11:59:31.8338 & 29:14:43.823 & 	bzq & 	LJET & 	0.729 & 	1250 & 	115.40 & 	7.84 & 	2.36e+12 & 	\nodata & 	\nodata	\\
J12030+6031 & 	1FGL J1202.9+6032 & CRATES J1203+6031 & 12:03:03.5094 & 60:31:19.129 & 	agn & 	LJET & 	0.065 & 	145 & 	44.79 & 	20.07 & 	2.75e+10 & 	\nodata & 	\nodata	\\
J12089+5441 & 	1FGL J1209.3+5444 & CRATES J1208+5441 & 12:08:54.2583 & 54:41:58.19 & 	bzq & 	PS & 	1.344 & 	252 & 	35.45 & 	15.20 & 	2.28e+11 & 	\nodata & 	\nodata	\\
J12093+4119 & 	1FGL J1209.4+4119 & B3 1206+416 & 12:09:22.7851 & 41:19:41.36 & 	bzb & 	LJET & 	0.377\tablenotemark{N} & 	131 & 	22.25 & 	9.57 & 	2.91e+10 & 	\nodata & 	\nodata	\\
J12098+1810 & 	1FGL J1209.7+1806 & CRATES J1209+1810 & 12:09:51.7649 & 18:10:06.796 & 	bzq & 	SJET & 	0.845 & 	140 & 	19.72 & 	7.97 & 	3.19e+10 & 	\nodata & 	\nodata	\\
J12178+3007 & 	1FGL J1217.7+3007 & B2 1215+30 & 12:17:52.0838 & 30:07:00.625 & 	bzb & 	LJET & 	0.13 & 	261 & 	83.01 & 	31.55 & 	7.36e+10 & 	\nodata & 	\nodata	\\
F12215+7106 & 	1FGL J1221.5+7106 & CRATES J1220+7105 & 12:20:03.631 & 71:05:31.09 & 	bzq & 	SJET & 	0.451 & 	183 & 	10.07 & 	3.84 & 	9.96e+10 & 	\nodata & 	\nodata	\\
J12201+3431 & 	1FGL J1220.2+3432 & CGRaBS J1220+3431 & 12:20:08.2902 & 34:31:21.711 & 	bzb & 	LJET & 	0.643\tablenotemark{N} & 	129 & 	11.03 & 	4.65 & 	1.90e+10 & 	\nodata & 	\nodata	\\
J12215+2813 & 	1FGL J1221.5+2814 & W Com & 12:21:31.6936 & 28:13:58.497 & 	bzb & 	LJET & 	0.102 & 	320 & 	76.04 & 	11.97 & 	3.64e+10 & 	\nodata & 	\nodata	\\
F12248+8044 & 	1FGL J1224.8+8044 & CRATES J1223+8040 & 12:23:40.500 & 80:40:04.30 & 	bzb & 	LJET & 	\nodata & 	450 & 	25.48 & 	10.12 & 	4.01e+10 & 	27 & 	-1	\\
J12248+4335 & 	1FGL J1225.8+4336\tablenotemark{*} & B3 1222+438 & 12:24:51.5074 & 43:35:19.276 & 	agu & 	LJET & 	1.07491\tablenotemark{N} & 	207 & 	36.15 & 	15.51 & 	3.15e+10 & 	\nodata & 	\nodata	\\
J12249+2122\tablenotemark{M} & 	1FGL J1224.7+2121 & 4C +21.35 & 12:24:54.4600 & 21:22:46.438 & 	bzq & 	LJET & 	0.435 & 	746 & 	69.70 & 	7.78 & 	3.97e+11 & 	81 & 	5	\\
J12269+4340 & 	1FGL J1225.8+4336\tablenotemark{*} & B3 1224+439 & 12:26:57.9051 & 43:40:58.438 & 	bzq & 	SJET & 	2.002 & 	85 & 	36.15 & 	15.51 & 	8.58e+9 & 	\nodata & 	\nodata	\\
J12302+2518 & 	1FGL J1230.4+2520 & ON 246 & 12:30:14.0935 & 25:18:07.145 & 	bzb & 	LJET & 	0.135 & 	206 & 	27.43 & 	11.46 & 	1.97e+10 & 	33 & 	-3	\\
F12316+2850 & 	1FGL J1231.6+2850 & B2 1229+29 & 12:31:43.579 & 28:47:49.81 & 	bzb & 	LJET & 	0.236 & 	86.3 & 	25.44 & 	8.13 & 	1.33e+10 & 	17 & 	7	\\
F12431+3627 & 	1FGL J1243.1+3627 & B2 1240+36 & 12:43:12.739 & 36:27:43.99 & 	bzb & 	SJET & 	1.0654\tablenotemark{N} & 	59.6 & 	24.30 & 	9.36 & 	8.77e+9 & 	\nodata & 	\nodata	\\
J12483+5820 & 	1FGL J1248.2+5820 & CGRaBS J1248+5820 & 12:48:18.784 & 58:20:28.725 & 	bzb & 	LJET & 	0.847\tablenotemark{N} & 	124 & 	61.70 & 	6.81 & 	3.85e+10 & 	36 & 	2	\\
J12531+5301 & 	1FGL J1253.0+5301 & CRATES J1253+5301 & 12:53:11.9232 & 53:01:11.741 & 	bzb & 	LJET & 	\nodata & 	240 & 	35.44 & 	6.32 & 	2.91e+10 & 	10 & 	3	\\
J12579+3229 & 	1FGL J1258.3+3227 & B2 1255+32 & 12:57:57.2313 & 32:29:29.321 & 	bzq & 	LJET & 	0.806 & 	467 & 	21.96 & 	0.00 & 	3.49e+10 & 	\nodata & 	\nodata	\\
J13030+2433 & 	1FGL J1303.0+2433 & CRATES J1303+2433 & 13:03:03.2143 & 24:33:55.684 & 	bzb & 	PS & 	0.993\tablenotemark{N} & 	134 & 	36.84 & 	6.88 & 	1.52e+11 & 	\nodata & 	\nodata	\\
F13060+7852 & 	1FGL J1306.0+7852 & CRATES J1305+7854 & 13:05:00.019 & 78:54:35.71 & 	agu & 	SJET & 	\nodata & 	213 & 	17.53 & 	6.82 & 	1.82e+10 & 	\nodata & 	\nodata	\\
J13083+3546 & 	1FGL J1308.5+3550 & CGRaBS J1308+3546 & 13:08:23.7095 & 35:46:37.16 & 	bzq & 	SJET & 	1.055 & 	309 & 	41.59 & 	16.49 & 	3.25e+11 & 	\nodata & 	\nodata	\\
F13092+1156 & 	1FGL J1309.2+1156 & 4C +12.46 & 13:09:33.931 & 11:54:24.59 & 	bzb & 	LJET & 	\nodata & 	629 & 	29.66 & 	12.32 & 	9.56e+9 & 	33 & 	8	\\
J13104+3220\tablenotemark{M} & 	1FGL J1310.6+3222 & B2 1308+32 & 13:10:28.6618 & 32:20:43.790 & 	bzq & 	LJET & 	0.997 & 	1240 & 	135.55 & 	8.98 & 	6.57e+11 & 	25 & 	-32	\\
J13127+4828 & 	1FGL J1312.4+4827 & CGRaBS J1312+4828 & 13:12:43.3508 & 48:28:30.928 & 	bzq & 	PS & 	0.501 & 	66.7 & 	31.29 & 	12.84 & 	1.62e+9 & 	\nodata & 	\nodata	\\
J13147+2348 & 	1FGL J1314.7+2346 & CRATES J1314+2348 & 13:14:43.8021 & 23:48:26.701 & 	bzb & 	LJET & 	\nodata & 	123 & 	35.58 & 	15.30 & 	2.22e+10 & 	\nodata & 	\nodata	\\
J13176+3425 & 	1FGL J1317.8+3425 & B2 1315+34A & 13:17:36.4935 & 34:25:15.921 & 	bzq & 	LJET & 	1.05 & 	260 & 	18.93 & 	7.43 & 	4.47e+10 & 	\nodata & 	\nodata	\\
J13211+2216 & 	1FGL J1321.1+2214 & CGRaBS J1321+2216 & 13:21:11.2041 & 22:16:12.098 & 	bzq & 	SJET & 	0.943 & 	244 & 	34.18 & 	14.92 & 	2.14e+10 & 	\nodata & 	\nodata	\\
F13213+8310 & 	1FGL J1321.3+8310 & CRATES J1321+8316 & 13:21:45.590 & 83:16:13.40 & 	agu & 	LJET & 	1.024\tablenotemark{N} & 	232 & 	27.88 & 	11.62 & 	1.32e+10 & 	9 & 	-9	\\
J13270+2210\tablenotemark{M} & 	1FGL J1326.6+2213 & B2 1324+22 & 13:27:00.8577 & 22:10:50.150 & 	bzq & 	SJET & 	1.4 & 	1120 & 	54.09 & 	11.78 & 	1.22e+12 & 	\nodata & 	\nodata	\\
J13307+5202 & 	1FGL J1331.0+5202 & CGRaBS J1330+5202 & 13:30:42.5962 & 52:02:15.448 & 	agn & 	LJET & 	0.688 & 	148 & 	41.52 & 	0.00 & 	4.50e+10 & 	\nodata & 	\nodata	\\
J13327+4722 & 	1FGL J1332.9+4728 & B3 1330+476 & 13:32:45.2413 & 47:22:22.653 & 	bzq & 	PS & 	0.669 & 	298 & 	32.08 & 	14.24 & 	1.30e+11 & 	\nodata & 	\nodata	\\
J13338+5057 & 	1FGL J1333.2+5056 & CLASS J1333+5057 & 13:33:53.7823 & 50:57:35.914 & 	agu & 	PS & 	1.362\tablenotemark{N} & 	48.4 & 	64.33 & 	27.63 & 	1.46e+10 & 	\nodata & 	\nodata	\\
J13455+4452 & 	1FGL J1345.4+4453 & B3 1343+451 & 13:45:33.1685 & 44:52:59.581 & 	bzq & 	SJET & 	2.534 & 	333 & 	52.18 & 	21.04 & 	2.11e+11 & 	\nodata & 	\nodata	\\
J13508+3034 & 	1FGL J1351.0+3035 & B2 1348+30B & 13:50:52.7333 & 30:34:53.582 & 	bzq & 	SJET & 	0.714 & 	251 & 	14.81 & 	5.25 & 	3.29e+10 & 	\nodata & 	\nodata	\\
F13533+1434 & 	1FGL J1353.3+1434 & PKS 1350+148 & 13:53:22.841 & 14:35:39.30 & 	bzb & 	LJET & 	\nodata & 	177 & 	24.77 & 	10.48 & 	9.44e+9 & 	32 & 	29	\\
F13581+7646 & 	1FGL J1358.1+7646 & CGRaBS J1357+7643 & 13:57:55.370 & 76:43:21.00 & 	bzq & 	SJET & 	1.585 & 	425 & 	40.70 & 	17.83 & 	8.58e+10 & 	\nodata & 	\nodata	\\
J13590+5544 & 	1FGL J1359.1+5539 & CRATES J1359+5544 & 13:59:05.7379 & 55:44:29.362 & 	bzq & 	PS & 	1.014 & 	108 & 	37.91 & 	16.61 & 	6.36e+10 & 	\nodata & 	\nodata	\\
J14270+2348 & 	1FGL J1426.9+2347 & PKS 1424+240 & 14:27:00.3942 & 23:48:00.045 & 	bzb & 	CPLX & 	\nodata & 	199 & 	69.51 & 	7.92 & 	1.21e+10 & 	\nodata & 	\nodata	\\
J14340+4203 & 	1FGL J1433.9+4204 & B3 1432+422 & 14:34:05.6956 & 42:03:16.01 & 	bzq & 	SJET & 	1.24 & 	131 & 	25.17 & 	10.63 & 	1.50e+10 & 	\nodata & 	\nodata	\\
J14366+2321 & 	1FGL J1436.9+2314 & PKS 1434+235 & 14:36:40.9873 & 23:21:03.297 & 	bzq & 	LJET & 	1.545 & 	600 & 	24.69 & 	10.78 & 	2.79e+11 & 	\nodata & 	\nodata	\\
J14388+3710 & 	1FGL J1438.7+3711\tablenotemark{*} & B2 1436+37B & 14:38:53.6095 & 37:10:35.408 & 	bzq & 	LJET & 	2.401 & 	263 & 	33.44 & 	15.00 & 	2.31e+11 & 	\nodata & 	\nodata	\\
F14387+3711 & 	1FGL J1438.7+3711\tablenotemark{*} & CRATES J1439+3712 & 14:39:20.580 & 37:12:02.81 & 	bzq & 	LJET & 	1.021 & 	33.3 & 	33.44 & 	15.00 & 	9.52e+9 & 	\nodata & 	\nodata	\\
F14438+2457 & 	1FGL J1443.8+2457 & PKS 1441+25 & 14:43:56.890 & 25:01:44.51 & 	bzq & 	LJET & 	0.939 & 	286 & 	20.05 & 	7.96 & 	1.15e+11 & 	\nodata & 	\nodata	\\
J14509+5201 & 	1FGL J1451.0+5204 & CLASS J1450+5201 & 14:50:59.9877 & 52:01:11.7 & 	bzb & 	PS & 	\nodata & 	37.5 & 	33.61 & 	14.31 & 	2.25e+10 & 	\nodata & 	\nodata	\\
J14544+5124 & 	1FGL J1454.6+5125 & CRATES J1454+5124 & 14:54:27.1247 & 51:24:33.734 & 	bzb & 	LJET & 	1.08\tablenotemark{N} & 	80.2 & 	50.54 & 	22.17 & 	5.83e+9 & 	\nodata & 	\nodata	\\
F15044+1029 & 	1FGL J1504.4+1029 & PKS 1502+106 & 15:04:24.979 & 10:29:39.19 & 	bzq & 	LJET & 	1.839 & 	1030 & 	1061.22 & 	14.67 & 	3.54e+10 & 	56 & 	-18	\\
J15061+3730 & 	1FGL J1505.8+3725 & B2 1504+37 & 15:06:09.5287 & 37:30:51.128 & 	bzq & 	LJET & 	0.674 & 	630 & 	31.13 & 	12.90 & 	8.54e+10 & 	11 & 	3	\\
J15169+1932 & 	1FGL J1516.9+1928 & PKS 1514+197 & 15:16:56.7985 & 19:32:13.01 & 	bzb & 	LJET & 	1.07\tablenotemark{N} & 	662 & 	35.06 & 	15.62 & 	2.27e+11 & 	\nodata & 	\nodata	\\
F15197+4216 & 	1FGL J1519.7+4216 & B3 1518+423 & 15:20:39.720 & 42:11:11.51 & 	bzq & 	PS & 	0.484 & 	40.8 & 	20.75 & 	8.72 & 	1.09e+10 & 	\nodata & 	\nodata	\\
J15221+3144 & 	1FGL J1522.1+3143 & B2 1520+31 & 15:22:09.9947 & 31:44:14.427 & 	bzq & 	CPLX & 	1.487 & 	462 & 	411.49 & 	10.60 & 	1.12e+11 & 	\nodata & 	\nodata	\\
J15396+2744 & 	1FGL J1539.7+2747 & CGRaBS J1539+2744 & 15:39:39.141 & 27:44:38.288 & 	bzq & 	SJET & 	2.19 & 	145 & 	13.50 & 	4.90 & 	9.70e+10 & 	\nodata & 	\nodata	\\
J15429+6129 & 	1FGL J1542.9+6129 & CRATES J1542+6129 & 15:42:56.9464 & 61:29:55.358 & 	bzb & 	LJET & 	\nodata & 	97.8 & 	65.53 & 	6.88 & 	2.38e+10 & 	\nodata & 	\nodata	\\
F15534+1255 & 	1FGL J1553.4+1255 & PKS 1551+130 & 15:53:32.700 & 12:56:51.68 & 	bzq & 	LJET & 	1.308 & 	658 & 	119.49 & 	13.66 & 	1.21e+10 & 	29 & 	-157	\\
F15557+1111 & 	1FGL J1555.7+1111 & PG 1553+113 & 15:55:43.039 & 11:11:24.40 & 	bzb & 	SJET & 	0.36 & 	160 & 	77.62 & 	22.73 & 	6.35e+10 & 	\nodata & 	\nodata	\\
J16046+5714 & 	1FGL J1604.3+5710 & CGRaBS J1604+5714 & 16:04:37.3568 & 57:14:36.668 & 	bzq & 	LJET & 	0.72 & 	374 & 	54.28 & 	23.42 & 	1.01e+10 & 	3 & 	14	\\
J16071+1551 & 	1FGL J1607.1+1552 & 4C +15.54 & 16:07:06.4276 & 15:51:34.495 & 	agn & 	LJET & 	0.496 & 	322 & 	33.90 & 	7.75 & 	3.19e+10 & 	27 & 	4	\\
F16090+1031 & 	1FGL J1609.0+1031 & 4C +10.45 & 16:08:46.200 & 10:29:07.80 & 	bzq & 	LJET & 	1.226 & 	1120 & 	61.34 & 	9.82 & 	4.12e+10 & 	\nodata & 	\nodata	\\
J16136+3412\tablenotemark{M} & 	1FGL J1613.5+3411 & B2 1611+34 & 16:13:41.0633 & 34:12:47.903 & 	bzq & 	LJET & 	1.397 & 	2960 & 	22.67 & 	9.88 & 	3.58e+10 & 	\nodata & 	\nodata	\\
J16160+4632 & 	1FGL J1616.1+4637 & CRATES J1616+4632 & 16:16:03.7689 & 46:32:25.239 & 	bzq & 	PS & 	0.95 & 	74.2 & 	37.69 & 	16.18 & 	4.16e+10 & 	\nodata & 	\nodata	\\
F16302+5220 & 	1FGL J1630.2+5220 & CRATES J1630+5221 & 16:30:43.150 & 52:21:38.59 & 	bzb & 	PS & 	\nodata & 	27.3 & 	19.32 & 	7.35 & 	1.59e+9 & 	\nodata & 	\nodata	\\
F16354+8228 & 	1FGL J1635.4+8228 & NGC 6251 & 16:32:31.980 & 82:32:16.40 & 	agn & 	LJET & 	0.025 & 	637 & 	45.51 & 	18.77 & 	1.41e+10 & 	1 & 	2	\\
J16377+4717 & 	1FGL J1637.9+4707 & 4C +47.44 & 16:37:45.1338 & 47:17:33.822 & 	bzq & 	LJET & 	0.74 & 	673 & 	35.84 & 	8.79 & 	7.56e+10 & 	\nodata & 	\nodata	\\
F16410+1143 & 	1FGL J1641.0+1143 & CRATES J1640+1144 & 16:40:58.889 & 11:44:04.20 & 	agn & 	LJET & 	0.078 & 	115 & 	47.36 & 	20.81 & 	7.65e+9 & 	\nodata & 	\nodata	\\
J16475+4950 & 	1FGL J1647.4+4948 & CGRaBS J1647+4950 & 16:47:34.9142 & 49:50:00.586 & 	agn & 	LJET & 	0.047 & 	154 & 	33.29 & 	8.54 & 	2.02e+10 & 	\nodata & 	\nodata	\\
J16568+6012 & 	1FGL J1656.9+6017 & CRATES J1656+6012 & 16:56:48.2475 & 60:12:16.455 & 	bzq & 	PS & 	0.623 & 	323 & 	18.12 & 	7.82 & 	7.16e+10 & 	\nodata & 	\nodata	\\
F17001+6830 & 	1FGL J1700.1+6830 & CGRaBS J1700+6830 & 17:00:09.300 & 68:30:07.02 & 	bzq & 	PS & 	0.301 & 	372 & 	55.99 & 	7.32 & 	7.28e+10 & 	\nodata & 	\nodata	\\
J17096+4318 & 	1FGL J1709.6+4320 & B3 1708+433 & 17:09:41.0876 & 43:18:44.547 & 	bzq & 	LJET & 	1.027 & 	140 & 	42.84 & 	17.10 & 	2.19e+10 & 	\nodata & 	\nodata	\\
F17192+1745 & 	1FGL J1719.2+1745 & PKS 1717+177 & 17:19:13.049 & 17:45:06.41 & 	bzb & 	SJET & 	0.137 & 	642 & 	44.76 & 	15.61 & 	1.66e+11 & 	\nodata & 	\nodata	\\
F17225+1012 & 	1FGL J1722.5+1012 & CRATES J1722+1013 & 17:22:44.580 & 10:13:35.80 & 	bzq & 	SJET & 	0.732 & 	386 & 	59.14 & 	26.13 & 	1.93e+10 & 	\nodata & 	\nodata	\\
J17240+4004 & 	1FGL J1724.0+4002 & B2 1722+40 & 17:24:05.4301 & 40:04:36.457 & 	agn & 	LJET & 	1.049 & 	435 & 	47.16 & 	8.97 & 	3.32e+11 & 	\nodata & 	\nodata	\\
F17250+1151 & 	1FGL J1725.0+1151 & CGRaBS J1725+1152 & 17:25:04.339 & 11:52:15.49 & 	bzb & 	SJET & 	0.018\tablenotemark{N} & 	63.3 & 	54.96 & 	23.25 & 	4.47e+10 & 	\nodata & 	\nodata	\\
J17274+4530 & 	1FGL J1727.3+4525 & B3 1726+455 & 17:27:27.6472 & 45:30:39.743 & 	bzq & 	SJET & 	0.714 & 	1050 & 	39.28 & 	7.93 & 	1.68e+12 & 	\nodata & 	\nodata	\\
J17283+5013 & 	1FGL J1727.9+5010 & I Zw187 & 17:28:18.6238 & 50:13:10.48 & 	bzb & 	LJET & 	0.055 & 	133 & 	28.67 & 	12.02 & 	6.88e+9 & 	11 & 	31	\\
F17308+3716 & 	1FGL J1730.8+3716 & CRATES J1730+3714 & 17:30:47.050 & 37:14:55.10 & 	bzb & 	SJET & 	\nodata & 	54.9 & 	24.08 & 	9.56 & 	1.91e+9 & 	\nodata & 	\nodata	\\
J17343+3857 & 	1FGL J1734.4+3859 & B2 1732+38A & 17:34:20.5821 & 38:57:51.446 & 	bzq & 	PS & 	0.976 & 	872 & 	99.48 & 	10.85 & 	1.22e+11 & 	\nodata & 	\nodata	\\
J17425+5945 & 	1FGL J1742.1+5947 & CRATES J1742+5945 & 17:42:32.0074 & 59:45:06.729 & 	bzb & 	LJET & 	\nodata & 	118 & 	27.81 & 	7.02 & 	1.58e+10 & 	1 & 	-1	\\
F17442+1934 & 	1FGL J1744.2+1934 & 1ES 1741+196 & 17:43:57.830 & 19:35:08.99 & 	bzb & 	SJET & 	0.083 & 	154 & 	15.54 & 	6.15 & 	1.12e+10 & 	\nodata & 	\nodata	\\
F17485+7004 & 	1FGL J1748.5+7004 & CGRaBS J1748+7005 & 17:48:32.839 & 70:05:50.78 & 	bzb & 	CPLX & 	0.77 & 	615 & 	31.49 & 	12.02 & 	2.88e+10 & 	\nodata & 	\nodata	\\
J17490+4321 & 	1FGL J1749.0+4323 & B3 1747+433 & 17:49:00.3604 & 43:21:51.287 & 	bzb & 	LJET & 	\nodata & 	373 & 	35.44 & 	14.39 & 	6.46e+10 & 	\nodata & 	\nodata	\\
F17566+5524 & 	1FGL J1756.6+5524\tablenotemark{*} & CRATES J1757+5523 & 17:57:28.279 & 55:23:11.90 & 	bzb & 	LJET & 	0.065 & 	40.7 & 	27.79 & 	12.00 & 	4.04e+9 & 	\nodata & 	\nodata	\\
F18004+7827 & 	1FGL J1800.4+7827 & CGRaBS J1800+7828 & 18:00:45.679 & 78:28:04.01 & 	bzb & 	LJET & 	0.68 & 	2370 & 	64.41 & 	7.67 & 	1.28e+11 & 	12 & 	-8	\\
F18070+6945 & 	1FGL J1807.0+6945 & 3C 371 & 18:06:50.681 & 69:49:28.09 & 	bzb & 	LJET & 	0.051 & 	1200 & 	75.93 & 	12.26 & 	4.46e+10 & 	6 & 	14	\\
F18096+2908 & 	1FGL J1809.6+2908 & CRATES J1809+2910 & 18:09:45.389 & 29:10:19.88 & 	bzb & 	SJET & 	\nodata & 	96.9 & 	21.61 & 	8.26 & 	1.32e+11 & 	\nodata & 	\nodata	\\
F18134+3141 & 	1FGL J1813.4+3141 & B2 1811+31 & 18:13:35.210 & 31:44:17.59 & 	bzb & 	LJET & 	0.117 & 	81.3 & 	30.48 & 	11.57 & 	1.31e+10 & 	\nodata & 	\nodata	\\
F18240+5651 & 	1FGL J1824.0+5651 & 4C +56.27 & 18:24:07.070 & 56:51:01.51 & 	bzb & 	LJET & 	0.664\tablenotemark{N} & 	1060 & 	64.59 & 	10.22 & 	1.82e+11 & 	\nodata & 	\nodata	\\
F18485+3224 & 	1FGL J1848.5+3224 & B2 1846+32A & 18:48:22.099 & 32:19:02.60 & 	bzq & 	LJET & 	0.798 & 	322 & 	74.19 & 	15.38 & 	3.55e+10 & 	\nodata & 	\nodata	\\
F18493+6705 & 	1FGL J1849.3+6705 & CGRaBS J1849+6705 & 18:49:16.080 & 67:05:41.71 & 	bzq & 	LJET & 	0.657 & 	1310 & 	227.47 & 	9.47 & 	3.84e+11 & 	\nodata & 	\nodata	\\
F18525+4853 & 	1FGL J1852.5+4853 & CGRaBS J1852+4855 & 18:52:28.550 & 48:55:47.50 & 	bzq & 	PS & 	1.25 & 	420 & 	51.36 & 	22.81 & 	7.41e+11 & 	\nodata & 	\nodata	\\
F19030+5539 & 	1FGL J1903.0+5539 & CRATES J1903+5540 & 19:03:11.611 & 55:40:38.39 & 	bzb & 	LJET & 	\nodata & 	180 & 	37.53 & 	15.16 & 	2.49e+10 & 	35 & 	-4	\\
F19416+7214 & 	1FGL J1941.6+7214 & CRATES J1941+7221 & 19:41:26.981 & 72:21:42.19 & 	agu & 	SJET & 	\nodata & 	157 & 	66.46 & 	29.70 & 	2.18e+10 & 	\nodata & 	\nodata	\\
F20000+6508 & 	1FGL J2000.0+6508 & 1ES 1959+650 & 19:59:59.849 & 65:08:54.71 & 	bzb & 	LJET & 	0.049 & 	213 & 	71.92 & 	8.60 & 	2.31e+10 & 	39 & 	2	\\
F20019+7040 & 	1FGL J2001.9+7040 & CRATES J2001+7040 & 20:01:33.950 & 70:40:25.82 & 	agu & 	LJET & 	\nodata & 	36.5 & 	27.18 & 	10.17 & 	5.63e+9 & 	\nodata & 	\nodata	\\
F20060+7751 & 	1FGL J2006.0+7751 & CGRaBS J2005+7752 & 20:05:31.001 & 77:52:43.21 & 	bzb & 	LJET & 	0.342 & 	904 & 	51.58 & 	22.83 & 	9.70e+10 & 	\nodata & 	\nodata	\\
F20091+7228 & 	1FGL J2009.1+7228 & 4C +72.28 & 20:09:52.301 & 72:29:19.39 & 	bzb & 	LJET & 	\nodata & 	775 & 	46.83 & 	13.78 & 	1.62e+10 & 	\nodata & 	\nodata	\\
F20204+7608 & 	1FGL J2020.4+7608 & CGRaBS J2022+7611 & 20:22:35.590 & 76:11:26.20 & 	bzb & 	LJET & 	\nodata & 	774 & 	45.59 & 	19.42 & 	1.19e+11 & 	33 & 	27	\\
F20315+1219 & 	1FGL J2031.5+1219 & PKS 2029+121 & 20:31:55.001 & 12:19:41.30 & 	bzb & 	SJET & 	1.215\tablenotemark{N} & 	1080 & 	50.42 & 	21.03 & 	4.74e+11 & 	\nodata & 	\nodata	\\
F20354+1100 & 	1FGL J2035.4+1100 & PKS 2032+107 & 20:35:22.339 & 10:56:06.79 & 	bzq & 	LJET & 	0.601 & 	522 & 	74.93 & 	14.31 & 	1.36e+11 & 	\nodata & 	\nodata	\\
F20497+1003 & 	1FGL J2049.7+1003\tablenotemark{*} & PKS 2047+098 & 20:49:45.859 & 10:03:14.40 & 	agu & 	LJET & 	\nodata & 	698 & 	58.69 & 	25.99 & 	6.07e+10 & 	\nodata & 	\nodata	\\
F21155+2937 & 	1FGL J2115.5+2937 & B2 2113+29 & 21:15:29.419 & 29:33:38.41 & 	bzq & 	LJET & 	1.514 & 	705 & 	39.17 & 	11.71 & 	7.26e+9 & 	\nodata & 	\nodata	\\
F21161+3338 & 	1FGL J2116.1+3338 & B2 2114+33 & 21:16:14.520 & 33:39:20.41 & 	bzb & 	LJET & 	\nodata & 	49.5 & 	38.92 & 	15.69 & 	4.46e+9 & 	\nodata & 	\nodata	\\
F21209+1901 & 	1FGL J2120.9+1901 & OX 131 & 21:21:00.610 & 19:01:28.31 & 	bzq & 	LJET & 	2.18 & 	363 & 	43.68 & 	17.31 & 	2.25e+10 & 	21 & 	5	\\
F21434+1742 & 	1FGL J2143.4+1742 & OX 169 & 21:43:35.539 & 17:43:48.68 & 	bzq & 	SJET & 	0.211 & 	728 & 	184.60 & 	10.49 & 	1.52e+11 & 	\nodata & 	\nodata	\\
F21525+1734 & 	1FGL J2152.5+1734 & PKS 2149+17 & 21:52:24.821 & 17:34:37.81 & 	bzb & 	SJET & 	0.871 & 	495 & 	20.25 & 	7.86 & 	1.51e+10 & 	\nodata & 	\nodata	\\
F21574+3129 & 	1FGL J2157.4+3129 & B2 2155+31 & 21:57:28.819 & 31:27:01.40 & 	bzq & 	SJET & 	1.486 & 	479 & 	53.99 & 	10.54 & 	3.57e+11 & 	\nodata & 	\nodata	\\
F22035+1726 & 	1FGL J2203.5+1726 & PKS 2201+171 & 22:03:26.890 & 17:25:48.29 & 	bzq & 	LJET & 	1.076 & 	692 & 	93.40 & 	8.59 & 	6.52e+10 & 	\nodata & 	\nodata	\\
F22121+2358 & 	1FGL J2212.1+2358 & PKS 2209+236 & 22:12:05.971 & 23:55:40.58 & 	bzq & 	LJET & 	1.125 & 	902 & 	21.37 & 	8.40 & 	1.30e+11 & 	\nodata & 	\nodata	\\
F22171+2423 & 	1FGL J2217.1+2423 & B2 2214+24B & 22:17:00.830 & 24:21:46.01 & 	bzb & 	LJET & 	0.505 & 	550 & 	42.39 & 	8.32 & 	3.30e+11 & 	\nodata & 	\nodata	\\
F22193+1804 & 	1FGL J2219.3+1804 & CGRaBS J2219+1806 & 22:19:14.090 & 18:06:35.60 & 	bzq & 	SJET & 	1.071 & 	169 & 	25.49 & 	10.40 & 	4.80e+10 & 	\nodata & 	\nodata	\\
F22362+2828 & 	1FGL J2236.2+2828 & B2 2234+28A & 22:36:22.469 & 28:28:57.40 & 	bzq & 	SJET & 	0.795 & 	1500 & 	84.96 & 	8.23 & 	2.62e+11 & 	\nodata & 	\nodata	\\
F22440+2021 & 	1FGL J2244.0+2021 & CRATES J2243+2021 & 22:43:54.739 & 20:21:03.82 & 	bzb & 	LJET & 	\nodata & 	60.5 & 	38.41 & 	14.31 & 	1.51e+10 & 	\nodata & 	\nodata	\\
F22501+3825 & 	1FGL J2250.1+3825 & B3 2247+381 & 22:50:05.750 & 38:24:37.19 & 	bzb & 	SJET & 	0.119 & 	56.5 & 	26.04 & 	10.72 & 	5.33e+9 & 	\nodata & 	\nodata	\\
F22517+4030 & 	1FGL J2251.7+4030 & CRATES J2251+4030 & 22:51:59.770 & 40:30:58.21 & 	bzb & 	SJET & 	\nodata & 	63.2 & 	46.51 & 	19.73 & 	9.07e+9 & 	\nodata & 	\nodata	\\
F22539+1608 & 	1FGL J2253.9+1608 & 3C 454.3 & 22:53:57.751 & 16:08:53.59 & 	bzq & 	LJET & 	0.859 & 	11200 & 	1355.47 & 	17.05 & 	5.91e+11 & 	36 & 	24	\\
F23073+1452 & 	1FGL J2307.3+1452 & CGRaBS J2307+1450 & 23:07:34.001 & 14:50:17.99 & 	bzb & 	LJET & 	0.503\tablenotemark{N} & 	50.7 & 	24.64 & 	9.21 & 	2.32e+10 & 	\nodata & 	\nodata	\\
F23110+3425 & 	1FGL J2311.0+3425 & B2 2308+34 & 23:11:05.330 & 34:25:10.88 & 	bzq & 	LJET & 	1.817 & 	808 & 	53.68 & 	9.52 & 	5.80e+10 & 	\nodata & 	\nodata	\\
F23220+3208 & 	1FGL J2322.0+3208 & B2 2319+31 & 23:21:54.950 & 32:04:07.61 & 	bzq & 	LJET & 	1.489 & 	475 & 	40.48 & 	17.44 & 	2.32e+11 & 	\nodata & 	\nodata	\\
F23216+2726 & 	1FGL J2321.6+2726 & 4C +27.50 & 23:21:59.861 & 27:32:46.39 & 	bzq & 	LJET & 	1.253 & 	941 & 	41.26 & 	18.15 & 	4.25e+10 & 	\nodata & 	\nodata	\\
F23226+3435 & 	1FGL J2322.6+3435 & CRATES J2322+3436 & 23:22:44.011 & 34:36:13.90 & 	bzb & 	PS & 	0.098 & 	27.3 & 	34.87 & 	0.00 & 	6.68e+9 & 	\nodata & 	\nodata	\\
F23252+3957 & 	1FGL J2325.2+3957 & B3 2322+396 & 23:25:17.870 & 39:57:36.50 & 	bzb & 	SJET & 	\nodata & 	139 & 	27.66 & 	8.30 & 	1.24e+10 & 	\nodata & 	\nodata	\\
\enddata
\tablecomments{Col.\ (1): Source name: if name starts with 'J' it is a VIPS or MOJAVE source; if name starts with 'F' it is a new source.
Col.\ (2): 1FGL source name.  
Col.\ (3): Alternate source name.
Col.\ (4): Right Ascension (J2000).
Col.\ (5): Declination (J2000).
Col.\ (6): Optical Type (1LAC): bzb = BL Lac object, bzq = FSRQ, agn = non-blazar AGN, agu = AGN of uncertain type, sbg = starburst galaxy.
Col.\ (7): Automated Source Classification: LJET = Long Jet, SJET = Short Jet, PS = Point Source, CPLX = Complex. 
Col.\ (8): Redshift, from 1LAC (Abdo et al. 2010b) except where specified.
Col.\ (9): Total VLBA flux density at 5 GHz.
Col.\ (10): $\gamma$-ray flux in units of $10^{-9}$ photons cm$^{-2}$ s$^{-1}$ for 100 MeV to 100 GeV.
Col.\ (11): Error in the $\gamma$-ray flux in the same units as the flux.
Col.\ (12): Core Brightness Temperature as measured by automated program (Helmboldt et al 2007).
Col.\ (13): Opening Angle in degrees.
Col.\ (14): Change in jet position angle in degrees.
}
\tablenotetext{M}{MOJAVE source. For more data, visit the MOJAVE website http://www.physics.purdue.edu/astro/MOJAVE/MOJAVEIItable.html or see Lister et al. (2009b)}
\tablenotetext{*}{Indicates a LAT source which is associated with multiple radio sources with high ($\geq$80\%) probability in 1LAC}
\tablenotetext{N}{Redshift obtained from NED}
\label{datatable}
\end{deluxetable}

\textwidth = 7.0truein
\textheight = 10.0truein
\begin{deluxetable}{ccrrrrrr}
\tablecolumns{8}
\tabletypesize{\scriptsize}
\tablewidth{0pt}
\tablecaption{Source Classifications}
\tablehead{
\colhead{LAT/non-LAT} & \colhead{Opt Type}	&	\colhead{LJET}	&	\colhead{SJET}	&	\colhead{PS}	&	\colhead{CPLX}	&	\colhead{CSO}	&	\colhead{Unidentified} \\}
\startdata
LAT-detected &  &  &  &  &  &  &  \\
 & BL Lacs & 55 (58\%) & 25 (26\%) & 12 (13\%) & 3 (3\%) & \nodata & \nodata \\
 & FSRQs & 54 (50\%) & 30 (28\%) & 21 (20\%) & 2 (2\%) & \nodata & \nodata \\
 & AGN/Other & 21 (70\%) & 5 (17\%) & 4 (13\%) & \nodata & \nodata & \nodata \\
non-LAT-detected &  &  &  &  &  &  &  \\
 & BL Lacs & 11 (46\%) & 7 (29\%) & 6 (25\%) & \nodata & \nodata & \nodata \\
 & FSRQs & 188 (39\%) & 121 (25\%) & 136 (28\%) & 2 ($\sim$1\%) & 30 (6\%) & 2 ($\sim$1\%) \\
 & AGN/Other & 214 (42\%) & 98 (19\%) & 111 (21\%) & 11 (2\%) & 71 (14\%) & 10 (2\%) \\
\enddata
\tablecomments{LJET = long jet, SJET = short jet, PS = point source, CPLX = complex, CSO = compact symmetric object candidate
}
\label{morphtable}
\end{deluxetable}

\clearpage
\textwidth = 7.2truein

\clearpage

\clearpage

\clearpage

\clearpage

\clearpage

\clearpage

\clearpage

\clearpage

\clearpage

\clearpage

\clearpage
\begin{figure}
\epsscale{0.8}
\plotone{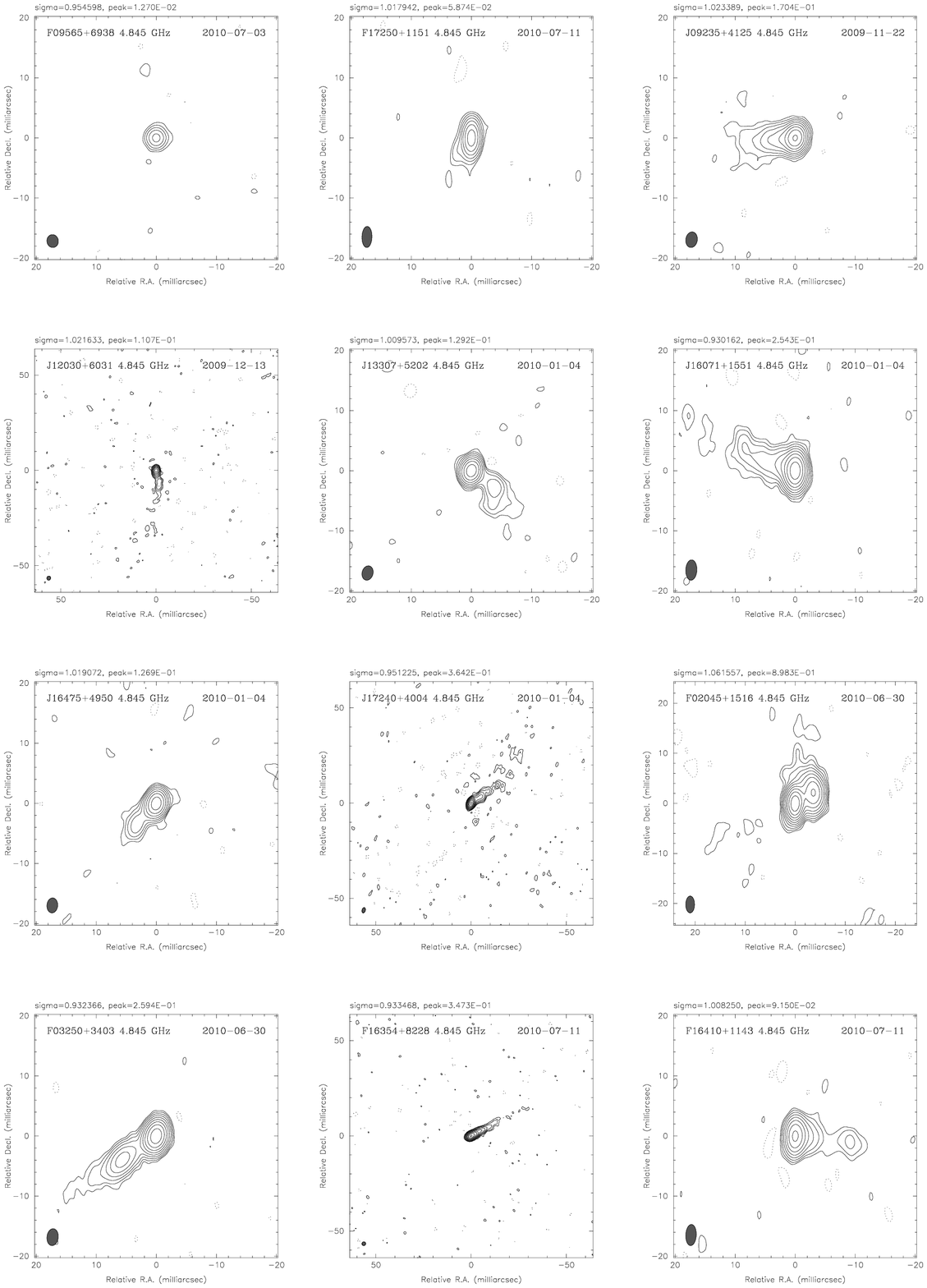}
\caption[A]{Contour maps of sources in Section 7.  All data were taken with the VLBA at 5 GHz.  Bottom contours are all at 0.6 mJy/beam, except for F02045+1516 which has a bottom contour of 1 mJy/beam.  The restoring beam is shown as an ellipse in the lower left corner.}
\label{imfig1}
\end{figure}

\clearpage
\begin{figure}
\ContinuedFloat
\epsscale{0.8}
\plotone{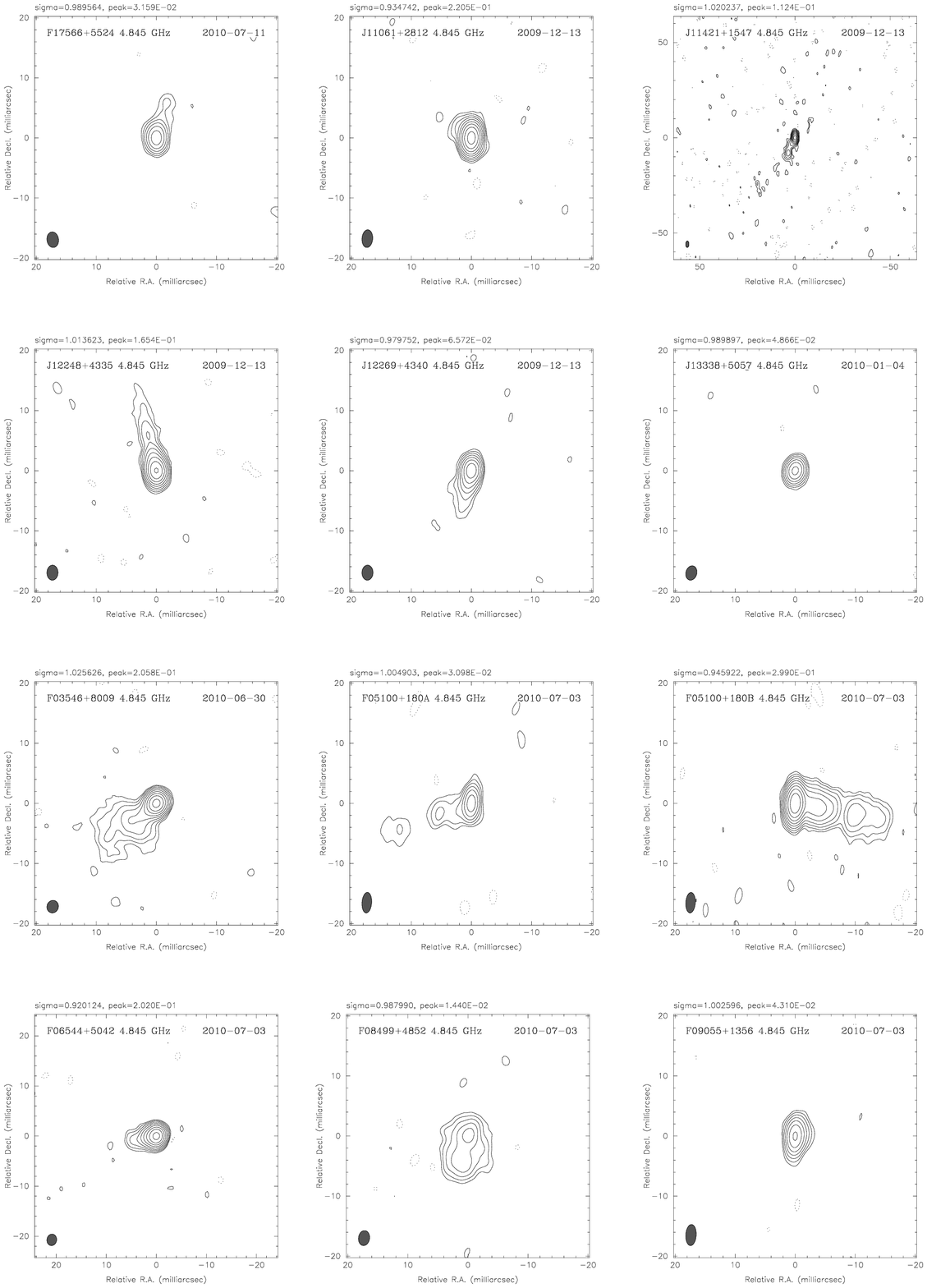}
\caption[B]{Contour maps of sources in Section 7.  All data were taken with the VLBA at 5 GHz.  Bottom contours are all at 0.6 mJy/beam.  The restoring beam is shown as an ellipse in the lower left corner.}
\label{imfig2}
\end{figure}

\clearpage
\begin{figure}
\ContinuedFloat
\epsscale{0.8}
\plotone{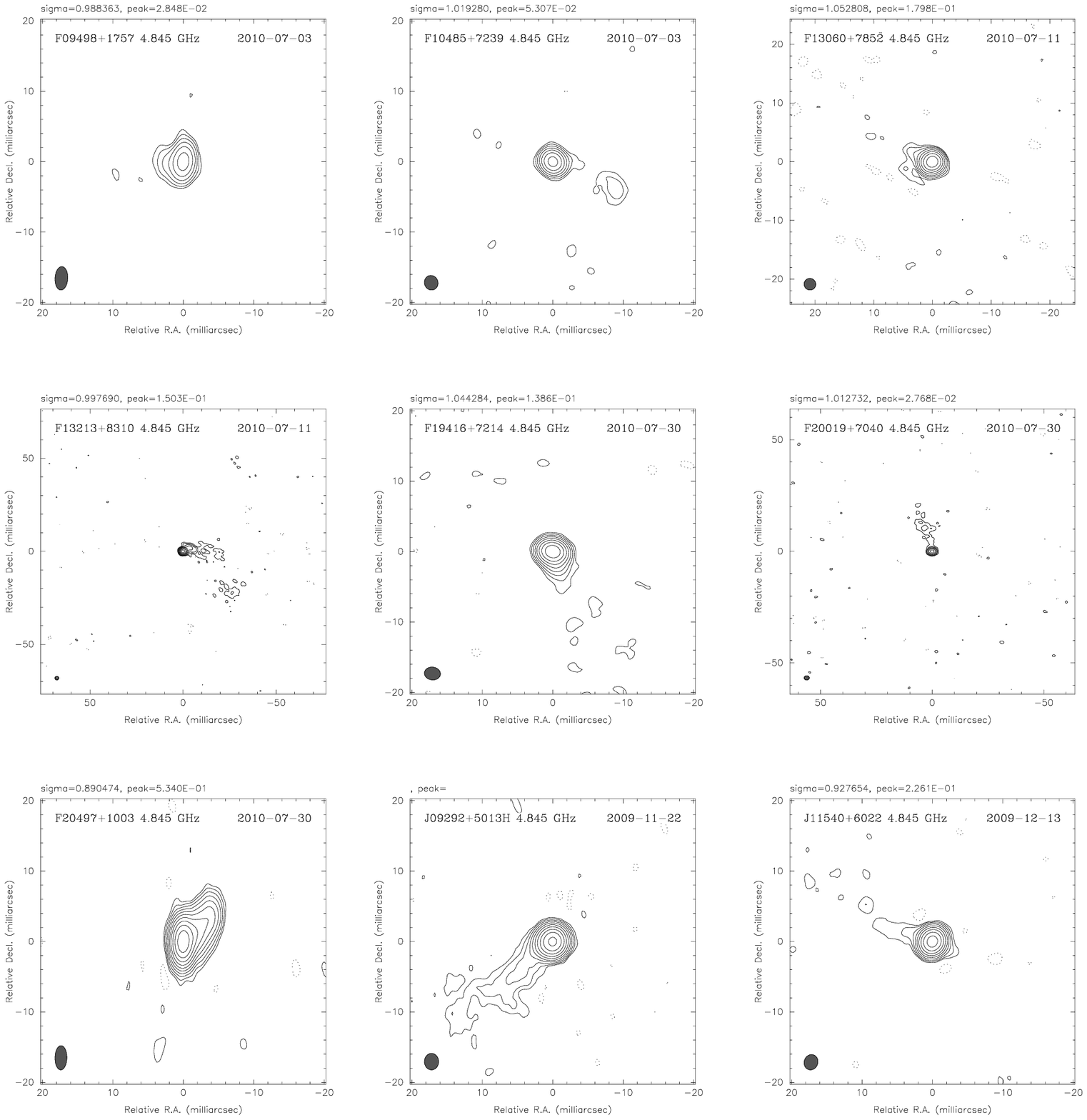}
\caption[C]{Contour maps of sources in Section 7.  All data were taken with the VLBA at 5 GHz.  Bottom contours are all at 0.6 mJy/beam, except for F13060+7852 and F13213+8310 which have bottom contours of 0.8 mJy/beam.  The restoring beam is shown as an ellipse in the lower left corner.}
\label{imfig3}
\end{figure}

\clearpage

\end{document}